\documentclass[twocolumn,
amsmath,amssymb,prx,
aps,longbibliography
]{revtex4-2}

\usepackage{graphicx}
\usepackage{dcolumn}
\usepackage{bm}
\usepackage{braket}
\usepackage{mathtools}
\usepackage{bbm}
\usepackage{comment}
\usepackage{makecell}
\usepackage{afterpage}
\usepackage{siunitx}
\usepackage{nicefrac}
\usepackage{amsthm}

\usepackage{hyperref}
\hypersetup{
	colorlinks=true,
	linkcolor=blue,
	citecolor=blue,
	filecolor=magenta,
	urlcolor=cyan
}

\def\iu{{\rm i}}

\DeclareMathOperator{\Tr}{Tr}
\DeclareMathOperator{\Var}{Var}
\DeclareMathOperator{\Cov}{Cov}
\def\dif{{\rm d}}

\newtheorem{theorem}{Theorem}

\newtheorem*{lemma*}{Lemma}

\newtheorem*{corollary*}{Corollary}

\graphicspath{{../}{../tikz/}}

\makeatletter
\DeclareFontFamily{OMX}{MnSymbolE}{}
\DeclareSymbolFont{MnLargeSymbols}{OMX}{MnSymbolE}{m}{n}
\SetSymbolFont{MnLargeSymbols}{bold}{OMX}{MnSymbolE}{b}{n}
\DeclareFontShape{OMX}{MnSymbolE}{m}{n}{
	<-6>  MnSymbolE5
	<6-7>  MnSymbolE6
	<7-8>  MnSymbolE7
	<8-9>  MnSymbolE8
	<9-10> MnSymbolE9
	<10-12> MnSymbolE10
	<12->   MnSymbolE12
}{}
\DeclareFontShape{OMX}{MnSymbolE}{b}{n}{
	<-6>  MnSymbolE-Bold5
	<6-7>  MnSymbolE-Bold6
	<7-8>  MnSymbolE-Bold7
	<8-9>  MnSymbolE-Bold8
	<9-10> MnSymbolE-Bold9
	<10-12> MnSymbolE-Bold10
	<12->   MnSymbolE-Bold12
}{}

\let\llangle\@undefined
\let\rrangle\@undefined
\DeclareMathDelimiter{\llangle}{\mathopen}%
{MnLargeSymbols}{'164}{MnLargeSymbols}{'164}
\DeclareMathDelimiter{\rrangle}{\mathclose}%
{MnLargeSymbols}{'171}{MnLargeSymbols}{'171}
\makeatother

\newcommand{\blockstyle}[1]{\boldsymbol{#1}}

\def\TCM{{T.C.M. Group, Cavendish Laboratory, JJ Thomson Avenue, Cambridge CB3 0HE, United Kingdom}}

\begin{document}
	
	\def\papertitle{{Postselection-free learning of measurement-induced quantum dynamics}}
	\def\authornames{{Max McGinley}}

	\title{\papertitle}
	\author{Max McGinley}
	\affiliation{\TCM}
	
	\begin{abstract}
		We address how one can empirically infer properties of quantum states generated by dynamics involving measurements. Our focus is on many-body settings where the number of measurements is extensive, making brute-force approaches based on postselection intractable due to their exponential sample complexity. We introduce a general-purpose scheme that can be used to infer any property of the post-measurement ensemble of states (e.g.~the average entanglement entropy, or frame potential) using a scalable number of experimental repetitions. We first identify a general class of `estimable properties' that can be directly extracted from experimental data. Then, based on empirical observations of such quantities, we show how one can indirectly infer information about any particular given non-estimable quantity of interest through classical post-processing. Our approach is based on an optimization task, where one asks what are the minimum and maximum values that the desired quantity could possibly take, while ensuring consistency with observations. The true value of this quantity must then lie within a feasible range between these extrema, resulting in two-sided bounds. Narrow feasible ranges can be obtained by using a classical simulation of the device to determine which estimable properties one should measure. Even in cases where this simulation is inaccurate, unambiguous information about the true value of a given quantity realised on the quantum device can be learned. As an immediate application, we show that our method can be used to verify the emergence of quantum state designs in experiments. We identify some fundamental obstructions that in some cases prevent sharp knowledge of a given quantity from being inferred, and discuss what can be learned in cases where classical simulation is too computationally demanding to be feasible. In particular, we prove that any observer who cannot perform a classical simulation cannot distinguish the output states from those sampled from a maximally structureless ensemble.
	\end{abstract}
	
	\maketitle
	
	\section{Introduction}
	
	In quantum mechanics, measurements serve both as a means to extract information about a system, and as a form of dynamics in of themselves. Not only does the outcome of a measurement provide information about the pre-measurement state, but also it is used to update one's knowledge of the post-measurement state, effectively leading to a stochastic `collapse' the wavefunction. This effect is central to a number of longstanding ideas in quantum information science, including active error correction \cite{Lidar2013} and measurement-based quantum computation \cite{Raussendorf2001,Raussendorf2003}. In recent years, a great deal of interest has emerged in the study of many-body quantum states that are generated by such dynamics, leading to the discovery of measurement-induced entanglement phase transitions \cite{Li2018,Skinner2019,Chan2019,Li2019,Szyniszewski2019,Gullans2020,Bao2020,Fan2021, Noel2022, Hoke2023}, emergent quantum state designs \cite{Ho2022,Claeys2022,Choi2023,Cotler2023,Lucas2023}, and protocols for generating long-ranged entangled states via non-unitary dynamics \cite{Piroli2021,Verresen2022,Tantivasadakarn2022,Lu2022,Iqbal2023,Fossfeig2023,Bao2024}.
	
	The probabilistic nature of quantum measurements makes probing such phenomena in experiment a considerable challenge. This is because the states of interest cannot be prepared deterministically; rather, in each repetition of the experiment, we will obtain a different randomly chosen outcome, and hence a different post-measurement state. Using conventional learning techniques, any property of a given quantum state can only be inferred through repeated preparation and measurement, which in this context would only be possible if we run the experiment sufficiently many times such that each state is realised on multiple occasions. Such a `postselection'-based approach has a sample complexity that is exponential in the number of measurements \cite{Choi2023,Koh2023}, which is infeasible for many-body systems.
	
	In this paper, we introduce a method by which properties of the post-measurement ensemble of states can be learned from experimental data, without suffering from the exponential cost of postselection. Although a given ensemble-averaged quantity of interest (e.g.~the mean entanglement entropy of the conditional states) may not be directly accessible in the sense one usually thinks of, we show that information about its value can still be inferred indirectly based on independent observations of certain other quantities, which we call `estimable properties'. These latter quantities are constructed such that they can be directly computed using data obtained from a scalable number of experimental repetitions.
	
\begin{figure*}
		\centering
		\includegraphics[scale=1]{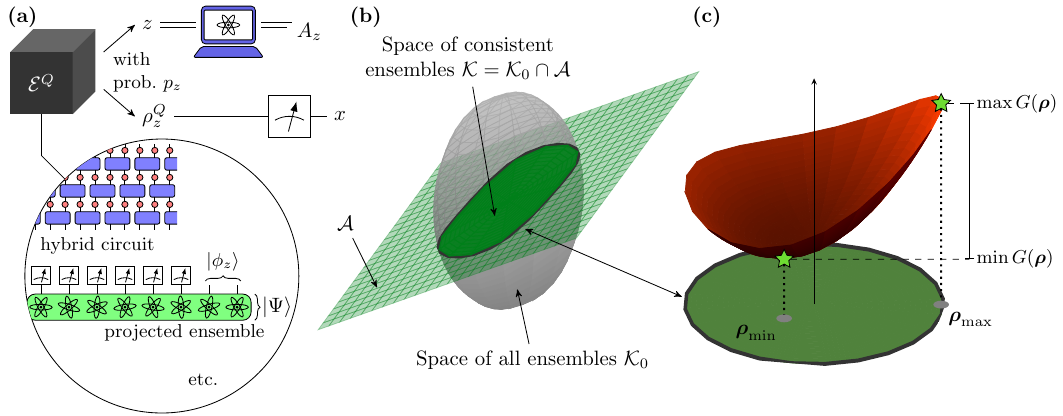}
		\caption{(a) In this work, we are interested in ensembles of quantum states $\mathcal{E}^Q$ [Eq.~\eqref{eq:Ensemble}] that arise from some quantum dynamics featuring measurement. This could be a hybrid quantum circuit, the projected ensemble, or some other protocol of interest. To be general, we can think of the device as a black box which on a given run of the experiment, with  with probability $p_z$ outputs a label $z$ (representing the outcome of all measurements during the dynamics), along with a corresponding quantum state $\rho^Q_z$ (the post-measurement conditional state). To learn properties of the ensemble, we perform a subsequent measurement on the state, here denoted $A_z$, which itself can depend on $z$; after many repetitions this allows us to infer properties of the form \eqref{eq:LinearObservableZ}. A classical simulation of the underlying dynamics can be used to help inform us how to best choose $A_z$ (Section \ref{sec:Simulations}). (b) Having learned the values of these measurable properties, we can characterise the space of all quantum state ensembles (collections of conditional states $\blockstyle{\rho} = \{\rho^Q_z\}$) that are consistent with our findings $\blockstyle{\rho} \in \mathcal{K}$---see Section \ref{sec:ConvOptimization}. (c) For a given ensemble property of interest $\bar{G}$ [Eq.~\eqref{eq:AverageProperty}], we can construct minimum and maximum values (green stars) over all ensembles the empirically deduced set $\mathcal{K}$. We can infer that the true value must lie between these extrema.}
		\label{fig:Main}
	\end{figure*}

	The basis of our method is to propose the following optimization task: What are the maximum and minimum values that the quantity of interest could take, based on the empirically observed values of a set of estimable properties? [See Fig.~\ref{fig:Main}(c).] These extrema provide us with two-sided bounds for the desired property, i.e.~we can conclude that the true value lies somewhere within this range. Despite the extremely high-dimensional nature of this optimization problem (scaling with the number of possible measurement outcomes), we show using analytical arguments how concrete bounds can be efficiently computed.
	
	The success of our scheme---as quantified by the gap between minimum and maximum---depends on which estimable properties we choose to measure. We propose to use classical simulations of the quantum device to inform this decision. That is, when we run the experiment and obtain a particular measurement outcome, a corresponding simulation of the dynamics is run on a classical device to determine what the best observable to measure is, given that the outcome in question has occurred. As the fidelity of the simulation improves, the bounds constructed using our method become tighter. We benchmark the performance of the method for particular representative examples, demonstrating that tight bounds can be constructed even when there is appreciable mismatch between the simulated and true states.
	
	This usage of a parallel simulation of the dynamics can be thought of as analogous to constructing `quantum-classical correlators', which have been recently introduced in the context of measurement-induced dynamics \cite{Li2021a,Barratt2022,Lee2022,Li2023,Feng2023,Garratt2023,Hoke2023} as an alternative to feedback-based approaches \cite{Gullans2020a, Noel2022}. In contrast to those works, where the inference one makes is dependent on how closely the quantum and classical devices behave, our approach allows one to extract unambiguous information regarding properties that are intrinsic to the quantum device alone. While bounds of this kind have been proved for specific quantities on a case-by-case basis in Ref.~\cite{Garratt2023a}, our method can in principle be used to estimate \textit{any} ensemble-averaged quantity, with a guarantee that the sharpness of the bounds is optimal for a given set of estimable parameters. We stress that even though model-based simulations are employed, the inferences we make are not contingent on any assumptions regarding the accuracy of this simulation, thereby allowing one to definitively verify whether or not some phenomena of interest is actually realised in experiment. As a concrete example, we show how the method introduced here can be used to verify the formation of quantum state designs in the projected ensemble \cite{Ho2022,Cotler2023}---a class of states with potential applications in state tomography, benchmarking, and cryptography \cite{Hayden2004,Knill2008, Radhakrishnan2009,Brandao2016, Mcginley2022a}.

	By analysing the nature of the information that one gains from performing experiments in general terms, we also identify certain fundamental limitations that in some cases preclude arbitrarily sharp knowledge about a certain property from being known, even in principle. In particular, we prove a result (Theorem \ref{thm:ExtremePoints}) which implies that if the conditional post-measurement states realised by the device are not close to being pure on average, then there will always be some residual uncertainty in the value of the desired quantity, even if the classical simulation is perfect. We also consider the possibility that classical simulations of the device may not be possible due to having too high a computational complexity. Our conclusion is that without the ability to perform some form of simulation, nothing can be learned about the ensemble of quantum states, besides the averaged state generated by the device (Theorem \ref{thm:Distinguish}). Put simply, this suggests that we should only think of the post-measurement conditional quantum states as being physically accessible if we have some means to predict something about their structure in advance. Altogether, our results establish a fundamental separation between what can and cannot be learned about measurement-induced dynamics, and pave the way towards an understanding of how the various phenomena that arise in this context can be leveraged for other purposes, be it quantum communication, cryptography, or computation.
	
	\section{Summary of main results}
	
	\subsection{Setup \label{subsec:Setup}}
	
	Our aim is to understand how one can infer properties of post-measurement quantum states from experimental data. Specifically, we consider settings where a quantum system $Q$ is subjected to some dynamics featuring measurements, and we are interested in properties of the conditional states of the system at the end of the dynamics. Scenarios that fit into this category include (but are not limited to):
	\begin{enumerate}
		\item Hybrid quantum circuits: models of dynamics typically defined in discrete time featuring measurements at various points in space and time \cite{Li2018,Skinner2019,Chan2019,Li2019,Szyniszewski2019,Bao2020,Fan2021}
		\item The projected ensemble, where a fixed many-body state is prepared before measurements of a subset of degrees of freedom are made \cite{Ho2022,Claeys2022,Bao2024,Choi2023,Cotler2023}
		\item Continuously monitored quantum systems \cite{Cao2019,Szyniszewski2019,Chen2020,Alberton2021,Li2021,Loio2023,Leontica2023}, and settings where one is interested in the quantum jump trajectories of open quantum systems \cite{Dalibard1992,Wiseman1993} 
		\item Systems undergoing error correction/detection, which involves measuring stabilizer operators \cite{Lidar2013}
		\item Measurement-based approaches to quantum computation \cite{Raussendorf2001,Raussendorf2003}
	\end{enumerate}
	Since measurements are fundamentally stochastic processes, the conditional state of the system after measurements have occurred is itself a random variable. Thus, in contrast to familiar scenarios where the task is to learn about a fixed quantum state $\rho$ that can be prepared deterministically, we are instead concerned with \textit{statistical ensembles} of quantum states
	\begin{align}
		\mathcal{E}^Q = \{(p_z, \rho^Q_z)\}_{z \in \mathcal{Z}}.
		\label{eq:Ensemble}
	\end{align}
	In the above, each  $z \in \mathcal{Z}$ is a possible measurement outcome, which occurs with probability $p_z$, and $\rho^Q_z$ is the (normalized) state of the system conditioned on this outcome---a unit-trace, Hermitian, positive semi-definite operator.

	Such ensembles of states can be thought of as an abstracted description of any of the above-mentioned examples. We will remain indifferent to the exact nature of the underlying dynamics, and instead picture a scenario where in each run of an experiment, some oracle (i.e.~a `black box', the implementation of which we disregard) provides us with a label $z$ (the measurement outcomes), sampled from the probability distribution $p_z$, along with a corresponding conditional state $\rho^Q_z$. In each run of the actual experiment, the dynamics of interest is executed, which we can think of as a single query to this oracle, after which we can apply some additional measurement whose purpose is to extract information about $\rho^Q_z$ [see Fig.~\ref{fig:Main}(a)].
	
	If the outcome $z$ is ignored, then this oracle is equivalent to a device that prepares the averaged state
	\begin{align}
		\braket{\rho^Q} \coloneqq \sum_z p_z \rho^Q_z
		\label{eq:DMAverage}
	\end{align}
	every time. Properties of $\braket{\rho^Q}$ can therefore be learned using conventional approaches. In some settings, we may be concerned with properties of this averaged state, while in others---including many of the examples listed above---the physics being investigated may be manifest in the individual conditional states that make up the ensemble \eqref{eq:Ensemble}. In the latter case, our wish is to learn properties of this ensemble $\mathcal{E}^Q$ beyond those of the average state, using some fixed number of queries/samples $M$.
	
	A well-appreciated issue that makes this objective difficult to achieve in the context of measurement-induced dynamics is the \textit{postselection problem}, which we describe in detail later. In brief, the problem stems from the fact that in each run of the experiment, we only get a single copy of the conditional state $\rho^Q_z$, which is sampled randomly from the distribution $p_z$. In the regime where the number of outcomes $|\mathcal{Z}|$ is large (which is to be expected in many-body settings), the state we get will be different every time for any reasonable number of repetitions $M$. Evidently, we can immediately rule out the possibility of learning properties of any individual conditional state $\rho^Q_{z^*}$, since the probability of this state never occurring is high. One might still hope to be able to estimate ensemble-averaged properties from a finite sample $\{z^{(1)}, \ldots, z^{(M)}\}$, i.e.~we look to estimate quantities of the form
	\begin{align}
		\mathbbm{E}_z[G] \coloneqq \sum_{z \in \mathcal{Z}} p_z G(\rho^Q_z),
		\label{eq:AverageProperty}
	\end{align}
	where $G(\sigma)$ is some function of a density matrix $\sigma$, and $\mathbbm{E}_z$ denotes an expectation values over the distribution $\{p_z\}$. However, in contrast to classical physics, quantum states cannot be copied, and thus having only single-copy access to the conditional states limits the information we can extract about each $\rho^Q_z$. In particular, as has been discussed previously, there is an apparent obstacle to learning properties of the kind \eqref{eq:AverageProperty} where $G(\sigma)$ is a nonlinear function of $\sigma$, since these cannot be expressed as functions of the average state $\braket{\rho^Q}$. A central aim of this paper is to critically examine this expectation, which is usually presented in somewhat heuristic terms, and to sharply determine exactly what can and cannot be inferred about a post-measurement quantum state ensemble from experimental data of a reasonable size $M$.
	
	\subsection{Results and structure of paper}
	
	Our first step to determine if and how the postselection problem can be circumvented is to establish a general class of ensemble properties that can be directly estimated using a reasonable number of repetitions $M$. In Section \ref{sec:Postselection}, we demonstrate that expressions of the form Eq.~\eqref{eq:LinearObservableZ} constitute such `estimable properties' of the ensemble, in that one can construct a function of the experimental data which equals the property in question on average, without any additional assumptions being made.
	
	Most properties of the ensemble that are of interest do not fall within this class, and hence cannot be directly estimated in the same way. Nevertheless, we demonstrate how information about some non-estimable property can be indirectly inferred using independent measurements of other estimable quantities, using the following logic. Given a collection of estimable quantities and some empirical observations of their values (which we get from running the experiment), we consider the space of all ensembles that are consistent with these observations. We refer to this as the feasible space $\mathcal{K}$ [Fig.~\ref{fig:Main}(b)], and we can guarantee that the true ensemble realised in the experiment lies somewhere within $\mathcal{K}$. We can characterise the best possible state of knowledge about some non-estimable average $\mathbbm{E}_z[G]$ by looking at the extremes of this quantity over the space $\mathcal{K}$. The maximum and minimum possible values of $\mathbbm{E}_z[G]$ that are consistent with our observations can be represented as the solutions to an \textit{optimization problem} [Fig.~\ref{fig:Main}(c)]. By solving these optimization problems, we can construct two-sided bounds for the desired quantity, i.e.~we infer that $\mathbbm{E}_z[G]$ must be between the minimum and maximum, both of which can be computed efficiently using a scalable number of repetitions $M$. Ideally the upper and lower bounds constructed using this approach will be close to one another, thereby giving us sharp knowledge about its value. This approach is outlined in detail in Section \ref{sec:ConvOptimization}, and we apply this idea to construct explicit bounds for various quantities of interest in Section \ref{sec:Constr}. 
	
	We are naturally concerned with how successful this indirect inference scheme can be, as quantified by the width of the feasible range. To address this issue, one must first decide how to choose which estimable parameters to measure. While our analysis works for any such choice, we can make a decision based on an approach introduced in previous works, where one constructs cross-correlations between experimental data and an independent simulation of the underlying dynamics. These `quantum-classical correlators', which we describe in Section \ref{sec:Simulations}, fall within the set of measurable properties, and hence can be efficiently estimated. (The nature of these simulations need not matter for the purposes of our inference scheme, but we address some specifics in the discussion.)
	
	Using this approach, we argue that the sharpness of the two-sided bounds depend on two key factors. Firstly, the accuracy of the simulation influences the gap between the minimum and maximum: Naturally, as the simulation gets closer to the true behaviour of the system, the bounds become tighter. Secondly, the nature of the conditional states that are realised on the quantum device $\rho^Q_z$ also plays an important role. We prove an important result---Theorem \ref{thm:ExtremePoints}---which states that regardless of which measurable quantities we compute, there will always be a consistent ensemble which is made up of states that are almost all pure. The implication is that when the actual states realised by the device $\rho^Q_z$ are mixed, we cannot necessarily constrain the range of a desired property $\mathbbm{E}_z[G]$ to be within an arbitrarily small window, even if the simulations we use are perfect. This represents a fundamental obstruction to learning mixed state ensembles without postselection, which we discuss in detail in Section \ref{subsec:MixedCons}.

	As an immediate application of our results, we show how the inference scheme developed here can be used to verify the emergence of quantum state designs in chaotic many-body systems \cite{Ho2022, Cotler2023}. We show how one can constrain a quantity known as the frame potential, which can be used to quantify how far an ensemble is from being a $k$-design (namely, an ensemble whose $k$th moments coincide with those of the Haar ensemble \cite{Ambainis2007}). In Section \ref{sec:Benchmarking}, we present a numerical simulation of an experiment that features both noise and miscalibrations in the Hamiltonian, and show that despite these imperfections (which are inevitably present in any experiment), the frame potential can be determined to be within a reasonably narrow window, in the sense describe above.
	
	In practice, performing a simulation of the quantum device may be a computationally demanding task, and we discuss the feasibility of such simulations for various specific cases in Section \ref{subsec:Computation}. It is therefore natural to consider whether anything can be learned in the case where simulation is not possible. To address this question, we present a result---Theorem \ref{thm:Distinguish}---the implication of which is that if we do not employ some sort of simulation of the dynamics (broadly defined), then we cannot learn anything about the ensemble of quantum states $\mathcal{E}^Q$ beyond properties of the averaged state \eqref{eq:DMAverage} using a number of repetitions that scales polynomially with system size. To be specific, in this regime the true ensemble is indistinguishable from one where every conditional state $\rho^Q_z$ is replaced by the averaged state \eqref{eq:DMAverage}. This suggests that an inability to simulate the device in question renders the conditional states inaccessible in experiment. We conclude by discussing this point, along with some of the other broader implications of our results in Section \ref{sec:Concl}.

	\section{Postselection problem: What can and can't be measured \label{sec:Postselection}}
	
	\subsection{The no-coincidence regime}
	
	In all the scenarios captured by our generalized setup, a natural task is to infer properties of the ensemble $\mathcal{E}^Q$ using some kind of learning scheme. In particular, for the purposes of this work we will be interested in estimating averaged properties of the states in the ensemble, i.e.~quantities that can be expressed in the form of Eq.~\eqref{eq:AverageProperty}.
	
	We start by addressing the postselection problem in detail. When it comes to learning properties of post-measurement quantum states from experimental data, a fundamental difficulty arises when the number of states in the ensemble $|\mathcal{Z}|$ is large---this is typically the case in the many-body setting, since $|\mathcal{Z}|$ scales exponentially with the number of measurements made, which itself typically scales with system size and/or time. This places us in a regime where, for any reasonable number of experimental repetitions $M$, the probabilities will satisfy $p_zM \ll 1$, meaning that we typically only get access to at most one instance of each state $\rho^Q_z$ over the whole experiment.
	
	The significance of this `no-coincidence' regime can be appreciated relatively straightforwardly: If we are given a single copy of a given quantum state $\rho^Q_z$, then whatever measurement we subsequently perform on it, the distribution of outcomes will depend linearly on the density matrix $\rho^Q_z$. Hence, if there are no coincidences (no value of $z$ occurs twice or more), then regardless of how we process the data, the only quantities that we can infer from the observed distribution of measurement outcomes are those that are themselves linear in $\rho^Q_z$. If we hastily apply this logic to quantities of the form \eqref{eq:AverageProperty}, this forces us to restrict ourselves to functions of the form $G(\rho^Q_z) = \Tr[\rho^Q_zA]$ for some observable $A$. In this case, we write the property in question as
	\begin{align}
		\braket{A}^Q \coloneqq \mathbbm{E}_z\Big[\!\Tr[\rho^Q_zA]\Big]  = \Tr[\braket{\rho^Q}A],
		\label{eq:ExpectationLinear}
	\end{align}
	where $\braket{\rho^Q}$ is the average state \eqref{eq:DMAverage}
	Such quantities evidently give us no information about the nature of individual states in the ensemble, and we only learn about the average state \eqref{eq:DMAverage}.
	
	In contrast, averages of nonlinear functions of the ensemble states, e.g.~squared expectation values $G(\rho^Q_z) = \Tr[A\rho^Q_z]^2$, do contain information beyond that contained in the average state, which is why these are the quantities that are of relevance to the various problems described in Section \ref{subsec:Setup}. If we had access to multiple copies of each state $\rho^Q_z$, then we could in principle learn such nonlinear functions by looking at the full distribution of measurement outcomes for each $z$ separately. However, in the regimes we are interested in this demands a prohibitively large number of experimental repetitions. Our aim is to find a solution to this postselection problem while keeping the query complexity bounded. 
	
	\subsection{Measurable quantities \label{subsec:Measurable}}
	
	To get a more precise picture of exactly what quantities are or are not experimentally accessible in the no-coincidence limit, let us consider a general procedure that can be used to extract information about the ensemble $\mathcal{E}^Q$. In a given repetition $r \in \{1, \ldots, M\}$, we obtain a sample from the ensemble $z^{(r)}$, and subsequently apply some (generalized) measurement to the quantum system $Q$, which can in principle depend on the outcome $z^{(r)}$. This $z$-dependent measurement scheme can be represented by a POVM $\mathcal{F}^Q(z) = \{F_x^Q(z) : x \in \mathcal{X}\}$, where $\mathcal{X}$ is a discrete set of measurement outcomes, and the operators $F_x^Q(z)$ are Hermitian positive semi-definite, satisfying $\sum_{x \in \mathcal{X}}F_x^Q(z) = \mathbbm{I}$ for each $z$. Conditioned on the outcome $z^{(r)}$, the result $x^{(r)} \in \mathcal{X}$ occurs with probability $\mathbbm{P}(x^{(r)}|z^{(r)}) = \Tr[F_{x^{(r)}}^Q(z^{(r)}) \cdot  \rho^Q_{z^{(r)}}]$, which is linear in $\rho^Q_{z^{(r)}}$, as discussed above, and together with the ensemble probabilities $p_z$ this defines a joint probability distribution for the set of possible outcomes of a single run
	\begin{align}
		\mathbbm{P}(x,z) = p_z \Tr[F_{x}^Q(z) \rho^Q_{z}].
		\label{eq:ExperimentalDistribution}
	\end{align}
	
	The full set of data that we acquire from the experiment $\{(x^{(r)}, z^{(r)}) : r = 1, \ldots M\}$ corresponds to a set of $M$ independent, identically distributed samples of pairs $X^{(r)} \coloneqq (x^{(r)},z^{(r)})$ drawn according to the probabilities \eqref{eq:ExperimentalDistribution}. Most obviously, from such a sample we can estimate the average of an arbitrary function of these pairs $f(X)$
	\begin{align}
		\overline{f(X)} \coloneqq \mathbbm{E}_X f(X) = \sum_{x,z} p_z \Tr[F^Q_x(z)\rho^Q_z] f(x,z),
		\label{eq:EstimableSingle}
	\end{align}
	where the overline is used as a shorthand for expectation values with respect to the samples $X$. The function $f(X^{(r)})$ is said to be an unbiased estimator of the quantity on the right hand side of \eqref{eq:EstimableSingle}. We will focus on the above quantities for the most part, since they are particularly relevant to ensemble averages of the form \eqref{eq:AverageProperty}. However, if we wish to be even more general, we could also use the sampled data to estimate functions of $n \leq M$ independently sampled pairs $f_n(X_1, \ldots, X_n)$
	\begin{align}
		\overline{f_n} &=\hspace*{-8pt} \sum_{\{x_i\}, \{z_i\}}  
		\hspace*{-2pt}\left[ \prod_{i=1}^np_{z_i}\!\Tr[F^Q_{x_i}(z_i)\rho^Q_{z_i}]\right]f_n\Big((x_1, z_1), \ldots, (x_n, z_n)\Big).
		\label{eq:EstimableMulti}
	\end{align}
	The quantities (\ref{eq:EstimableSingle}, \ref{eq:EstimableMulti}) are referred to as \textit{estimable parameters} of the distribution \eqref{eq:ExperimentalDistribution}, because one can find a function of the sampled data $\{X^{(1)}, \ldots, X^{(M)}\}$ that is equal to these quantities in expectation \cite{Bose2018, Ferguson2003}. In fact, any functional over the space of probability distributions that has an unbiased estimator must be expressible in the forms written above \cite{Halmos1946,Bickel1969}. Hence, these are the only classes of observables that we can experimentally learn in the no-coincidence limit.
	
	Returning to Eq.~\eqref{eq:EstimableSingle}, we remark that the ensemble states $\rho^Q_z$ only appear through the outcome probabilities \eqref{eq:ExperimentalDistribution}, which as discussed in the previous section are linear in the density matrices. It is therefore helpful to rewrite the average \eqref{eq:EstimableSingle} as 
	\begin{align}
		\overline{f(X)} = \braket{A_z}^Q \coloneqq \sum_z p_z \Tr[A_z \rho^Q_z],
		\label{eq:LinearObservableZ}
	\end{align}
	where we define the family of operators
	\begin{align}
		A_z \coloneqq \sum_x F_x^Q(z)  f(x,z).
		\label{eq:OperatorFamilyF}
	\end{align}
	Eq.~\eqref{eq:LinearObservableZ} gives us a succinct characterization of the class of quantities that can be learned experimentally without prohibitive postselection overheads. (The more general estimators \eqref{eq:EstimableMulti} can always be decomposed in terms of the above.) We immediately see that Eq.~\eqref{eq:ExpectationLinear} is the special case where we disregard the classical information $z$ when choosing the measurement scheme and post-processing function $f$, i.e.~$f(x,z) = f(x)$ and $F_x^Q(z) = F_x^Q$, such that $A_z$ becomes $z$-independent. Therefore, quite naturally, we conclude that in order to probe properties of the ensemble that cannot be characterized solely by the average density matrix \eqref{eq:DMAverage}, we must adopt a learning strategy that itself depends explicitly on $z$. Notably, to do so necessarily requires us to have some \textit{a priori} knowledge about the relationship between the labels $z$ and the states $\rho^Q_z$. The essence of our scheme, which we describe in more detail in the following sections, is to use idealised classical computer simulations of the quantum device to inform us as to how $A_z$ should depend on $z$.

	\subsection{Aside: Avoiding mid-circuit measurement and feed-forward using classical shadows\label{subsec:Shadows}}
	
	As written, the measurable quantities \eqref{eq:LinearObservableZ} appear to require a feed-forward mechanism, where the outcome $z$ is used to decide what physical measurement to perform. Before describing our postselection-free inference scheme in detail, we briefly pause to explain how this aspect of the measurement procedure, which may be hard to implement in practice, can be avoided using ideas from classical shadow tomography \cite{Huang2020}. A similar approach has been outlined in Ref.~\cite{Garratt2023a}.
	
	Without feed-forward, we must fix the POVM $\mathcal{F}^Q(z)$ to be $z$-independent. We will make the key assumption that this fixed POVM is \textit{informationally complete} \cite{Scott2006}, i.e.~the collection of operators $\{F_x^Q\}_{x \in \mathcal{X}}$ span the full space of operators over $\mathcal{H}^Q$. In this case, one can find a (not necessarily unique) complementary set of operators $\tilde{F}_x^Q$ which overall have the following property
	\begin{align}
		\sum_x \Tr[F_x^Q \rho] \tilde{F}_x^Q &= \rho & \forall \rho \in \mathcal{B}(\mathcal{H}^Q)
		\label{eq:DualFrame}
	\end{align}
	Using the nomenclature of Ref.~\cite{Scott2006}, the POVM operators $F^Q_x$ constitute an operator frame, while $\tilde{F}^Q_x$ is the corresponding dual frame. With this construction in hand, any observable of the form \eqref{eq:LinearObservableZ} can be estimated from a set of experimentally measured data $\{(x^{(r)}, z^{(r)}) : r = 1, \ldots M\}$ by using the post-processing function
	\begin{align}
		f(x^{(r)}, z^{(r)}) = \Tr[A_{z^{(r)}} \tilde{F}^Q_{x^{(r)}}].   
		\label{eq:EstimatorAZ}
	\end{align}
	Combining Eqs.~(\ref{eq:OperatorFamilyF}, \ref{eq:DualFrame}), we see that the average of such a function equals the right hand side of \eqref{eq:LinearObservableZ} in expectation
	\begin{align}
		\overline{f(x,z)} = \braket{A_z}^Q.
	\end{align}
	The above prescription \ref{eq:EstimatorAZ} gives us an explicit way of computing an unbiased estimator for any quantity of the form \eqref{eq:LinearObservableZ}. Notably, we can do this without deciding on the operators $A_z$ in advance of the physical measurement of the system, i.e.~we can ``measure first, ask questions later'' \cite{Elben2023}. This feature of informationally complete POVMs means we do not need to employ adaptive schemes, where the physical measurement procedure is decided based on the sample $z$.
	
	A particularly straightforward way to implement an informationally complete POVM is to use classical shadow tomography \cite{Paini2019,Huang2020}, which was proposed as a useful way to study measurement-induced dynamics in Ref.~\cite{Garratt2023a}. In each run of the experiment, we apply a randomly chosen unitary $U_c$ from a pre-chosen ensemble $\mathcal{U} = \{(q_c, U_c)\}_{c \in \mathcal{C}}$, where $\mathcal{C}$ is an arbitrary discrete set, before performing a projective measurement in a fixed basis $\{\ket{m}\bra{m}\}$. For certain choices of $\mathcal{U}$, informational completeness is guaranteed, and the dual frame can be computed efficiently.
	
\begin{figure}
		\centering
		\includegraphics{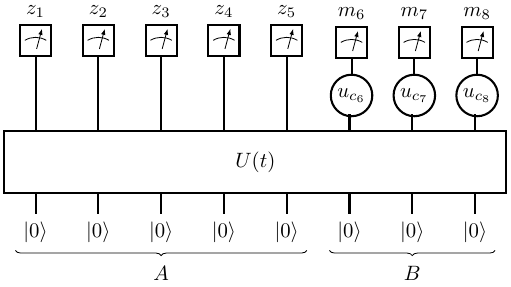}
		\caption{Protocol for measuring arbitrary estimable properties \eqref{eq:LinearObservableZ} for the projected ensemble using classical shadows, with all measurements occurring simultaneously at the end of the circuit. The projected ensemble corresponds to the collection of post-measurement states that arise when the bipartite state $\ket{\Psi^{AB}} = U(t)\ket{0^{\otimes N}}$ is prepared, and then qubits in $A$ are measured in the computational basis, with outcomes $\{z_i : i = 1,\ldots, N_A\}$. The post-measurement states are thus defined on the qubits in $B$; this figure shows the case $N_A = 5$, $N_B = 3$. As described in the main text, a shadow-based scheme can be used to probe these states, which corresponds to performing random on-site unitaries $u_{c_i}$ before measuring in the computational basis, with outcomes $m_i$ (to be distinguished from $z_i$, which label the states we are trying to probe). The dual frame \eqref{eq:DualFramePauli} is used when reconstructing estimable properties via Eq.~\eqref{eq:EstimatorAZ}.}
		\label{fig:shadow}
	\end{figure}

	While there are many such possibilities, as a concrete example, once can take $\mathcal{U}$ to be a uniform distribution over all single-qubit Clifford gates (i.e.~operations of the form $U_c = u_{c_1} \otimes \cdots \otimes u_{c_N}$, with $u_{c_i} \in \{I, H_X, H_Y\}$, with $H$, $H_Y$ Hadamard and $Y$-Hadamard gates, respectively). Then we can use the dual frame
	\begin{align}
		\tilde{F}_{(m,c)}^Q = \bigotimes_{i=1}^{N_Q}\big(3u_{c_i}^\dagger\ket{m_i}\bra{m_i}u_{c_i} - I\big).
		\label{eq:DualFramePauli}
	\end{align}
	Thus, the necessary measurement scheme can be implemented using single-qubit rotations and measurements. In many settings, this means we can perform all the measurements simultaneously at the end of the experiment (both those that generate the ensemble $\mathcal{E}^Q$ and those we use to extract information). This is illustrated for the case of the projected ensemble of a bipartite state $\ket{\Psi^{AB}} = U(t)\ket{0^{\otimes N}}$ in Fig.~\ref{fig:shadow}.
	
	The number of repetitions $M$ required to estimate $\braket{A_z}^Q$ to a given accuracy can be expressed in terms of the variance $\Var_{x,z}f(x,z) = \sum_z p_z \Tr[F^Q_x\rho^Q_z] \Tr[\tilde{F}^Q_x A_z]^2$. In the more familiar setting where one wishes to learning the properties of a fixed state $\rho$, the variance of a shadow tomographic estimator can be bounded using a useful construction the shadow norm $\|A\|_{\rm shadow}^2$, which is a function of the observable to be estimated $A$, as well as the ensemble $\mathcal{U}$; this is defined in Ref.~\cite{Huang2020}. In the present case, where we are instead dealing with an ensemble of states, it is straightforward to show that $\Var_{x,z}f(x,z)$ can be bounded as
	\begin{align}
		\Var_{x,z}f(x,z) \leq \sum_z p_z \|A_z\|_{\rm shadow}^2
	\end{align}
	Thus, the property $\braket{A_z}^Q$ can be estimated to a good accuracy using a reasonable number experimental repetitions provided that each of the operators $A_z$ is (with high probability) an operator that itself can be efficiently estimated in ordinary shadow tomography.
	
	The randomized Pauli measurements discussed above constitute one particular example of an informationally complete POVM, but we emphasise that there are many other alternative, e.g.~those based on global Clifford rotations \cite{Huang2020}, or even generalized methods that use chaotic Hamiltonian evolution and/or ancilla qubits \cite{Tran2023,Mcginley2022a}. In all the subsequent analysis, we remain agnostic to the exact measurement scheme used, and will simply assume that we have some way to measure the properties \eqref{eq:LinearObservableZ}.

	\section{Classical simulations \label{sec:Simulations}}
	
	We have now identified an explicit scheme for measuring quantities of the form $\braket{A_z}^Q$ [Eq.~\eqref{eq:LinearObservableZ}] without problematic postselection overheads. However, as mentioned previously, our goal is to infer properties of the ensemble that take the form \eqref{eq:AverageProperty}---specifically those for which $G(\rho)$ is a nonlinear function. Here we describe the idea of `quantum-classical correlators', where a simulations of the system in question is run in conjunction with the experimental, and describe what they can tell us about such nonlinear averages.\\

	By `classical simulations', we mean the following: Each time we perform a run of the experiment, which gives us a label $z \in \mathcal{Z}$ and a measurement outcome $x \in \mathcal{X}$, we also compute and store a representation of some corresponding state $\rho^C_z$ on a classical computer. We use the superscript $C$ to distinguish this `classical' state, which represents the result of some idealised simulation of the experiment, from the `quantum' state that is actually realised on the true device. We emphasise that this simulation can be done `lazily', i.e.~we only compute $\rho^C_z$ for those values of $z$ that happen to arise in the experiment, as opposed to pre-computing every state $\rho^C_z$ in advance, the cost of which would scale with $|\mathcal{Z}|$. We also do not need to classically sample from the distribution $p_z$, which in many cases is itself computationally demanding. The nature, accuracy, and computational cost of the classical simulation may depend on the specific physical setting being considered, and we will discuss several particular examples in Section \ref{subsec:Computation}. However, for the time being, we presume that some form of simulation is possible, and motivate our discussion on the basis that there is some partial correlation between the classical and quantum states, $\rho^C_z$ and $\rho^Q_z$. The case where a classical simulation is impossible---either due to incomplete knowledge of how the quantum device operates, or prohibitively high computational cost---is discussed in Section \ref{subsec:NoSim}.
	
	Let us take the simplest nontrivial case and suppose that our aim is to learn the average of a particular squared expectation value $\Tr[O \rho^Q_z]^2$ over the ensemble $\mathcal{E}^Q$. We introduce the following notation for such a quantity
	\begin{align}
		\braket{O\otimes O}^{QQ} &\coloneqq \sum_z p_z \Tr\Big[O^{\otimes 2} \cdot (\rho^Q_z)^{\otimes 2}\Big] \nonumber\\ &= \sum_z p_z \Tr\big[O\rho^Q_z\big]^2
		\label{eq:CorrQQ}
	\end{align}
	The superscript $QQ$ is used to emphasise that the quantity in question is a linear functional of the state $\rho^Q_z \otimes \rho^Q_z$. As explained in Section \ref{sec:Postselection}, we cannot directly measure this quantity. However, a natural proxy that has been introduced in several recent works is the `quantum-classical correlator' \cite{Li2021a,Barratt2022,Lee2022,Li2023,Feng2023,Garratt2023,Hoke2023}, which here we define as
	\begin{align}
		\braket{O\otimes O}^{QC} & \coloneqq \sum_z p_z \Tr[O\rho^Q_z]\Tr[O\rho^C_z ].
		\label{eq:CorrQC}
	\end{align}
	Note that the probabilities $p_z$ appearing in Eq.~\eqref{eq:CorrQC} are the same as those appearing in the fully quantum expression \eqref{eq:CorrQQ}. The above quantity can then be cast in the form of Eq.~\eqref{eq:LinearObservableZ} with
	\begin{align}
		A_z = \Tr[O\rho^C_z] \times O.
		\label{eq:AzCQ}
	\end{align}
	Thus, once we collect samples $z^{(r)}$ taken on the quantum device, we can compute the corresponding classical observables $\Tr[O\rho^C_z]$ and construct an estimator of the quantum-classical correlator, using the methods prescribed in Section \ref{sec:Postselection}.
	Finally, we can also consider a `classical-classical' correlator
	\begin{align}
		\braket{O\otimes O}^{CC} = \sum_z p_z \Tr\big[O\rho^C_z\big]^2
		\label{eq:CorrCC}
	\end{align}
	defined by analogy to Eq.~\eqref{eq:CorrQQ}, again with the probabilities $p_z$ set by the quantum device. This is also of the form \eqref{eq:LinearObservableZ}, with $A_z = \Tr[O\rho^C_z]^2 \times I$ (the quantum states are disregarded here).
	
	Clearly, in the limit where the classical simulation perfectly matches the behaviour of the quantum device, $\rho^C_z = \rho^Q_z$, all these quantities are equal. Thus, we hope that if the simulation is good, but not perfect, the proxy quantity \eqref{eq:CorrQC} will be close in value to the true `quantum-quantum' observable \eqref{eq:CorrQQ}, which is the physically relevant quantity. However, at present we cannot make any definitive conclusions about the value of the quantum-quantum correlator without making unsubstantiated assumptions about the accuracy of our classical simulations. Our objective in the following two sections, which form the most technical parts of this paper, is to establish methods that allow one to construct rigorous two-sided bounds for the true value \eqref{eq:CorrQQ} based on experimental observations of the measurable quantities (\ref{eq:CorrQC}, \ref{eq:CorrCC}), \textit{without} making any \textit{a priori} assumptions about how accurate the classical simulation is. That is, even though we are using our classical simulation as a form of prior `guess' for the conditional states $\rho^Q_z$, we allow for the possibility that this guess is incorrect. This skeptical approach to learning means that anything conclusions we make about the ensemble will be entirely unambiguous. 
	
	\section{Convex optimization approach to inferring averages\label{sec:ConvOptimization}}
	
	We have seen in detail how the no-coincidence limit gives rise to a distinction between properties of quantum state ensembles that can be measured straightforwardly---those of the form $\braket{A_z}_{\mathcal{E}^Q}$, Eq.~\eqref{eq:LinearObservableZ}, which include `quantum-classical' correlators \eqref{eq:CorrQC}---versus those that cannot be directly measured with a reasonable number of experimental repetitions, e.g.~the `quantum-quantum' correlator \eqref{eq:CorrQQ}. In the absence of direct estimation schemes for the latter class, we are interested in understanding the best possible state of knowledge that we could in principle have about such unobservable quantities, based on experimentally accessible data. Our intuition, based on the discussion of the previous section, is that by cross-correlating experimental outputs with classical simulations of the quantum system, we can gain some amount of knowledge about these quantities, even though we cannot measure them directly. To make this intuition concrete, in this and the following sections, we aim to address the question: Given knowledge of a set of observable quantities of the form \eqref{eq:LinearObservableZ}, what range of values can a particular unobservable quantity take, while ensuring consistency with our observations?
	
	More formally, suppose that from a set of experimental data, we construct estimates of a family of $R$ scalar quantities $\{\braket{A^{(i)}_z} : i = 1, \ldots, R\}$, the outcomes of which we denote $b_i$. For the moment, we presume that any statistical uncertainty in these observations can be neglected (an assumption which we will relax later). We wish to determine the maximum and minimum possible values that a particular average $\mathbbm{E}_z G(\rho^Q_z)$ can take over all quantum state ensembles $\mathcal{E}^Q$ that satisfy
	\begin{align}
		\label{eq:Consistent}
		\braket{A^{(i)}_z}_{\mathcal{E}^Q} &= b_i & \forall i \in \{1, \ldots, R\}.
	\end{align}
	Denoting the minimum and maximum values of $\mathbbm{E}_z G(\rho^Q_z)$ as $g_{\pm}^*$, this will give us a two-sided bound for the desired average, namely
	\begin{align}
		g_-^* \leq \mathbbm{E}_z G(\rho^Q_z) \leq g_+^*.
		\label{eq:FeasibleRange}
	\end{align}
	The determination of $g_{\pm}^*$ can be viewed as an \textit{optimization task}, where the object being varied over is the ensemble $\mathcal{E}^Q$ itself, and the function being extremized is the average $\mathbbm{E}_z G(\rho^Q_z)$. In this section, we study the structure of such optimization problems at a general level.
	
	\subsection{Set of consistent ensembles}
	
	Our first step is to analyse the structure of the space of quantum state ensembles that satisfy Eq.~\eqref{eq:Consistent}. For succinctness of notation, we will find it useful to represent the collection of states $\{\rho^Q_z\}_{z \in \mathcal{Z}}$ in terms of a single large block-diagonal matrix $\blockstyle{\rho} = \bigoplus_{z\in \mathcal{Z}}\rho^Q_z$, where each block contains the density matrix $\rho^Q_z$ for a particular label $z$. We can then view $\blockstyle{\rho}$ as an element of the linear space of matrices $\mathcal{M} \coloneqq \mathcal{B}(\mathcal{H})^{\oplus |\mathcal{Z}|}$. As for the probabilities $p_z$, while these are not known to us in full in practice, our ability to run the experiment means we can sample from this distribution; therefore our approach will to keep $p_z$ fixed, while allowing the states $\rho^Q_z$ themselves to vary. With all this in mind, we begin by formally specifying the space of valid quantum state ensembles as
	\begin{align}
		\mathcal{K}_0 \coloneqq \Big\{\blockstyle{\rho} = {\textstyle \bigoplus_{z \in \mathcal{Z}}}\rho^Q_z \Big|  \rho^Q_z \in \mathcal{D}\, \forall z \in \mathcal{Z}\Big\}
	\end{align}
	where $\mathcal{D} \subset \mathcal{B}(\mathcal{H})$ denotes the space of density matrices for a single copy of the system Hilbert space $\mathcal{H}$.
	
	We are interested specifically in ensembles that are consistent with the observations \eqref{eq:Consistent}. Again using the direct sum representation, for each block-diagonal matrix $\blockstyle{X} = \bigoplus_z X_z \in \mathcal{M}$, we define the linear function $A^{(i)}(\blockstyle{X}) \coloneqq \sum_z p_z \Tr[ A^{(i)}_z X_z]$, and we define the space $\mathcal{A}$ as
	\begin{align}
		\mathcal{A} \coloneqq \big\{ \blockstyle{X} \in \mathcal{M} \big| A^{(i)}(\blockstyle{X}) = b_i\, \forall\, i = 1, \ldots, R \big\}
	\end{align}
	Since $\mathcal{A}$ is a subspace of a linear space defined by $R$ linear constraints, we have that $\mathcal{A}$ is a hyperplane of codimension $R$ in $\mathcal{M}$.
	
	Finally, the \textit{feasible space} is given by the intersection $\mathcal{K} = \mathcal{K}_0 \cap \mathcal{A}$. Writing this out in full,
	\begin{align}
		\label{eq:FeasibleSet}
		\mathcal{K} \coloneqq \Big\{\blockstyle{\rho} = {\textstyle \bigoplus_{z}}\rho^Q_z \Big|\rho^Q_z \in \mathcal{D}\, \forall z \in \mathcal{Z} ;\, \mathbbm{E}_z \big(\!\Tr[A^{(i)}_z\rho^Q_z]\big) = b_i\Big\}.
	\end{align}
	Each element of $\mathcal{K}$ corresponds to a particular ensemble that could describe the system, based on the empirical data \eqref{eq:Consistent}.
	
	Crucially, we note that when viewed as a subset of the linear space $\mathcal{M}$, the set $\mathcal{K}$ is convex, i.e.~if $\blockstyle{\rho} = \bigoplus_z \rho^Q_z$ and $\blockstyle{\sigma} =  \bigoplus_z \sigma^Q_z$ are two sets of states that both belong to $\mathcal{K}$, then so too does
	\begin{align}
		\lambda\blockstyle{\rho}+ (1-\lambda)\blockstyle{\sigma} = \bigoplus_{z \in \mathcal{Z}} \big[\lambda \rho^Q_z + (1-\lambda)\sigma^Q_z \big]\in \mathcal{K}
	\end{align}
	for any $\lambda \in [0,1]$. This is a consequence of the convexity of both the space of density matrices $\mathcal{D}$ and the hyperplane $\mathcal{A}$, along with the fact that the intersection of two convex sets is itself convex.

	\subsection{Convex functions}
	
	Concretely, our aim is to determine the range of \textit{feasible values} that a particular average $\mathbbm{E}_z G$ can take over the space of feasible ensembles $\mathcal{K}$. Namely, we wish to characterize the set
	\begin{align}
		\mathcal{G} = \Big\{\, g \,\Big| \exists \blockstyle{\rho} \in \mathcal{K} \text{ such that }\bar{G}(\blockstyle{\rho}) = g\Big\},
	\end{align}
	where we introduce the shorthand
	\begin{align}
		\label{eq:AverageGBar}
		\bar{G}(\blockstyle{\rho}) \coloneqq \sum_z p_z G(\rho^Q_z).
	\end{align}
	The convexity of $\mathcal{K}$ will prove useful for this purpose, in particular for determining the extremal feasible values $g_+^* = \max_{g \in \mathcal{G}} g$ (similar for the minimum $g_-^*$). Indeed, we can formulate such a task in terms of the following optimization problems (which we write out in longhand momentarily)
	\begin{align}
		\label{eq:Optimize}
		g_\pm^* \coloneqq 
		\displaystyle \mathop{\text{max/min}}_{\bigoplus_z \rho^Q_z \in \mathcal{K}} \Bigg[\sum_z p_z G(\rho^Q_z)\Bigg],
	\end{align}
	(with $g_+^*$ corresponding to the maximum, and $g_-^*$ to the minimum). One can then immediately use these extrema to bound the average for the true ensemble $\mathcal{E}^Q$ on both sides as
	\begin{align}
		g_-^* \leq \bar{G}(\mathcal{E}^Q) \leq g_+^*.
	\end{align}
	This reflects the best possible state of knowledge we could have about the average $\bar{G}$, based on our observations.

	Optimization problems over convex sets such as Eq.~\eqref{eq:Optimize} have been well-studied in a wide variety of contexts. Most prominently, much work has gone into the study of \textit{convex optimization}, which is concerned with minimizing convex functions (equivalently, maximizing concave functions) over convex sets. Recall that a function $\bar{G} : \mathcal{K} \rightarrow \mathbbm{R}$ is convex if for any two $\blockstyle{\rho}, \blockstyle{\rho}' \in \mathcal{K}$ and any $\lambda \in [0,1]$, we have
	\begin{align}
		\bar{G}\big(\lambda\blockstyle{\rho} + (1-\lambda)\blockstyle{\rho}'\big) \leq \lambda \bar{G}(\blockstyle{\rho}) + (1-\lambda)\bar{G}(\blockstyle{\rho}')
	\end{align}
	Convex optimization tasks enjoy many useful properties, which can be exploited to gain analytical insight into their solutions, and to design efficient algorithms.
	For this reason, from hereon we specialize to cases where the function $\bar{G}(\blockstyle{\rho})$ is convex, unless mentioned otherwise. From our construction of $\bar{G}$ [Eq.~\eqref{eq:AverageGBar}], we see that $\bar{G}$ is convex if the  function $G(\rho^Q_z)$, the average of which we are interested in, is itself a convex function over density matrices. Some particularly pertinent functions that arise in quantum many-body physics include:
	
	\textit{1. Powers of expectation values.---}While expectation values of observables $\braket{O} = \Tr[\rho O]$ are linear in $\rho$, and hence directly estimable without any additional analysis, often we are interested in integer powers of expectation values, $\Tr[\rho O]^k$. The quantum-quantum correlator \eqref{eq:CorrQQ} is an example of this with $k = 2$. We would need to compute such a quantity arises if we wanted to know the variance of $\braket{O}$ over the ensemble $\mathcal{E}^Q$. We note that $\Tr[\rho O]^k$ is convex for even $k \geq 2$, or for any positive integer when $O$ is semi-positive definite.
	
	\textit{2. Entropies.---}Often we are interested in an entropy associated with a quantum state $\rho$, or a entropy of a subsystem of $Q$. The von Neumann entropy $S(\rho) = -\Tr[\rho\log \rho]$ is a concave function of $\rho$, as are R{\'e}nyi entropies $S_\alpha(\rho) = (1-\alpha)^{-1}\log(\Tr[\rho^\alpha])$ for $\alpha \in (0,1)$ \cite{Rastegin2011}. Since our subsequent analysis refers explicitly to convex functions, one can simply work with $-S(\rho)$, which is convex. One should then bear in mind that the role of minimization and maximimzation of the objective function are exchanged.
	
	\textit{3. Purities.---}A closely related object is the purity $\Tr[\rho^2]$ and generalizations to higher powers $\Tr[\rho^k]$. These are equal to exponentials of R{\'e}nyi entropies $S_{\alpha = k}(\rho)$, and hence give information on how mixed the states are, either globally or within a subsystem. It is straightforward to show that $\Tr[\rho^k]$ is convex in $\rho$ for $k \geq 1$.\\
	
	While our specialization to convex functions may seem restrictive, we note that any function $G(\rho)$ whose Hessian (matrix of second derivatives) is bounded can be decomposed as a sum of a convex and a concave function \cite{Yuille2001}, and hence each component can be bounded separately using the methods described in the following.
	
	\subsection{Duality and certificates for convex optimization problems}
	
	Working with the understanding that $\bar{G}(\blockstyle{\rho})$ is a convex function, we now describe our approach to finding the extrema $g_{\pm}^*$, or approximations thereof. Several standard techniques and results from the field of convex optimization will be used in the following; we refer the interested reader to Ref.~\cite{Boyd2004} for an introduction to the field and proofs of various pertinent results.

	The standard approach to optimization problems with equality constraints of the form \eqref{eq:Consistent} is to make use of Lagrange multipliers. For each constraint $i = 1, \ldots, R$ we introduce a scalar Lagrange multiplier $\lambda_i$, and define the Lagrange dual functions as follows \cite{Boyd2004}
	\begin{subequations}
		\begin{align}
			h_-(\lambda_i) &= \inf_{\blockstyle{\rho} \in \mathcal{K}_0} L(\blockstyle{\rho}, \lambda_i) \label{eq:DualFunctionMin} \\
			h_+(\lambda_i) &= \sup_{\blockstyle{\rho} \in \mathcal{K}_0} L(\blockstyle{\rho}, \lambda_i) 
			\label{eq:DualFunctionMax}
		\end{align}
		\label{eq:DualFunction}
	\end{subequations}
	where the Lagrangian $L(\blockstyle{\rho}, \lambda_i)$ is given by
	\begin{align}
		L(\blockstyle{\rho}, \lambda_i) \coloneqq \bar{G}(\blockstyle{\rho}) - \sum_i \lambda_i\big[A^{(i)}(\blockstyle{\rho}) - b_i\big]
	\end{align}
	We emphasise that the domains in Eqs.~\eqref{eq:DualFunction} is $\mathcal{K}_0$, without the linear constraints \eqref{eq:Consistent}. This is significant because $L(\blockstyle{\rho}, \lambda_i)$ is a sum over $z$, and thanks to the structure of $\mathcal{K}_0$, we can perform the extremization for each block $\rho^Q_z$ separately.
	
	An important concept in the study of optimization problems is that of strong and weak duality, which relate the original optimization task (the `primal problem') to a particular secondary task (the `dual problem'). The dual problem is to find the extreme points
	\begin{align}
		h^*_- &\coloneqq \max_{\lambda_i} h_-(\lambda_i),  & h^*_+ &\coloneqq \min_{\lambda_i} h_+(\lambda_i).
		\label{eq:CertificateIneqs}
	\end{align}
	Since $h_{\pm}(\lambda_i)$ are by definition the pointwise supremum and infimum of a family of affine functions, they are concave and convex functions, respectively. The dual problems are therefore convex optimization problems, even if the primal is not.
	
	Strong duality is the statement that the solutions to the primal and dual problems are identical. In our case, by Slater's condition \cite{Boyd2004}, strong duality holds for the minimization problem, i.e.~$h_-^*=g_-^*$, provided that $\bar{G}(\blockstyle{\rho})$ is convex. As for the maximization problem, this is not a convex optimization problem, since it is equivalent to minimizing $-\bar{G}$, which is concave by assumption. There is still some useful structure exhibited by convex maximization problems, which we will discuss in Section \ref{subsec:MaximizationExtreme}, but for the time being we will instead rely on weak duality, which holds irrespective of the nature of $\bar{G}$. Weak duality states that $h_-^* \leq g_-^*$ and $h_+^* \geq g_+^*$. This general property can be proved using the min-max inequality.
	
	While many efficient algorithms exist that allow one to solve convex optimization problems numerically, the high dimensionality of the primal problem (as well as the stochastic nature of the observations we make) mean that we cannot directly employ such algorithms as they stand. One option is to apply such methods to solve the dual problem, which has dimensionality $R$, rather than $|\mathcal{Z}|\times d^2$, thus making the problem more manageable. While this numerical approach is perfectly feasible, here we choose to study these problems analytically, in order to gain more insight into the structure of these problems. While exact solutions to the optimization problems are not always obtainable using analytic methods, we can still invoke strong or weak duality, which allow us to find \textit{primal-dual certificates}: That is, even if we only have an approximate solution $\{\lambda_i\}$ to the dual problem, we always have the following series of inequalities
	\begin{align}
		h_-(\lambda_i) &\leq h_-^* = g_-^*, & h_+(\lambda_i) &\geq h_+^* \geq g_+^*.
	\end{align}
	Thanks to these relations, if we can evaluate $h_-(\lambda_i)$ for some set of Lagrange multipliers, we can certify that the true minimum is no less than $h_-(\lambda_i)$. Thus, even if we cannot determine the exact value of the extrema, we can use suboptimal solutions of the dual problem to obtain conservative estimates of $g_\pm^*$, i.e.~a rigorous bound that is guaranteed \textit{not} to be an overestimate (underestimate) of the minimum (maximum). In lieu of an exact solution to either the primal or dual problems, our aim is then to find a way of obtaining $h_\pm(\lambda_i)$ for near-optimal $\lambda_i$, so as to obtain as tight a bound as possible.
	
	By treating these optimization problems analytically, rather than numerically, we will gain valuable insight that informs us how to choose the operators $A_z^{(i)}$ from the very beginning so as to obtain small feasible ranges; 
	we will have this in mind in Section \ref{sec:Constr} and beyond.
	
	\subsection{Minimization vs.~Maximization \label{subsec:MaximizationExtreme}}
	
	Before we begin the process of constructing explicit bounds for specific observables, it is important to point out a fundamental difference between minimization vs.~maximization problems for a given convex average $\bar{G}(\blockstyle{\rho})$. Namely, the former is a convex optimization problem (minimizing a convex function over a convex set), while the latter is not. This difference between the two will turn out to have important consequences for the tightness of the bounds that one can infer based on experimental observations.
	
	While the maximization problem cannot be cast into a standard convex optimzation problem, we can still invoke a useful property that results from its special structure: The maximum of a convex function over a convex set is always attained at at least one extreme point of the set. Recall that an extreme point of a convex set $\mathcal{C}$ are those elements $\tau \in \mathcal{C}$ which cannot be expressed as a nontrivial convex combination of two other elements. Specifically, $\tau$ is extreme if and only if $\tau = \lambda \tau' + (1-\lambda)\tau''$ for some $\lambda \in (0,1)$ implies $\tau' = \tau'' = \tau$ \cite{Boyd2004}. If $G(\tau)$ is a convex function over $\mathcal{C}$, then for any non-extreme $\tau$, we can find appropriate elements $\tau',\tau'' \neq \tau$ such that $G(\tau) = G(\lambda \tau' + (1-\lambda)\tau'') \leq \lambda G(\tau') + (1-\lambda)G(\tau'') \leq \max[G(\tau'),G(\tau'')]$, which proves the claim stated above.
	
	By considering the eigenstate decomposition of a density matrix $\rho$, one can see that the extreme points of the space $\mathcal{D}$ are pure states $\rho = \ket{\phi}\bra{\phi}$. This structure is naturally reflected in the extreme points of $\mathcal{K}$. In Appendix \ref{app:Extreme}, we prove the following:
	\begin{theorem}
		\label{thm:ExtremePoints}
		Any ensemble $\blockstyle{\rho}$ that is an extreme point of the set $\mathcal{K}$ [Eq.~\eqref{eq:FeasibleSet}] has at most $R$ states that are not globally pure, where $R$ is the number of linear constraints in Eq.~\eqref{eq:Consistent}.
	\end{theorem}
	Since $\mathcal{K}$ is necessarily nonempty (the true ensemble $\mathcal{E}^Q$ lies within $\mathcal{K}$), and is a closed, linearly bounded subset of $\mathcal{M} = \mathcal{B}(\mathcal{H})^{\oplus |\mathcal{Z}|}$, at least one extreme point must exist, and hence we immediately have
	\begin{corollary*}
		\label{corr:PureStateExist}
		Given the values of $R$ scalar measurable properties \eqref{eq:Consistent}, there exists an ensemble consistent with these observations for which at most $R$ states are globally non-pure.
	\end{corollary*}

	As a result, in the no-coincidence limit, where $R p_z \ll 1$, we can infer that the solution to the maximization problem is extremely close to the solution of the same problem with the additional constraint that all states are pure. This fact imposes fundamental limitations on how wide the feasible range \eqref{eq:FeasibleRange} can be made in cases where the actual state ensemble $\mathcal{E}^Q$ is significantly mixed, $\sum_z p_z \Tr[(\rho_z^Q)^2] < 1$. We will consider this issue in more detail in Section \ref{subsec:MixedCons}.

	\section{Constructing two-sided bounds \label{sec:Constr}}
	
	\bgroup
	\def\arraystretch{1.7}
	\begin{table*}
		\begin{ruledtabular}
			\begin{tabular}{lcccc}
				Quantity & Expression & Concave/convex? & Lower bound & Upper bound \\ \hline
				Global purity & $\mathbbm{E}_z\big[\!\Tr[(\rho^Q_z)^2]\big]$ & Convex & Eqs.~(\ref{eq:PurityMinimumL}, \ref{eq:PurityLBSuperoperator}) & Eq.~\eqref{eq:PurityUB}$^*$ \\ \hline
				\begin{minipage}{90pt}
					\raggedright
					Subsystem purity/\\quadratic observable
				\end{minipage}&
				\begin{minipage}{120pt}
					$\mathbbm{E}_z\big[\Tr[\rho^Q_z O]^2\big]$ \\
					or $\mathbbm{E}_z\big[\!\Tr[\rho^Q_z \mathcal{N}(\rho^Q_z)]\big],\; \mathcal{N} \succeq 0$
				\end{minipage}  & Convex & Eq.~\eqref{eq:QuadLower} & Eq.~\eqref{eq:QuadUpper} \\ \hline
				von Neumann entropy & $\mathbbm{E}_z\big[\!-\!\Tr[\rho^{Q_1}_z\log \rho^{Q_1}_z]\big]$ & Concave & Eqs.~(\ref{eq:vNMinInfNorm}, \ref{eq:vNMinSimple}) & Eqs.~(\ref{eq:vNMaxLambda}, \ref{eq:vNMaxQC}) \\ \hline
				Frame potential & $\mathbbm{E}_{z,z'} \Tr\big[\rho^Q_{z\vphantom{'}} \rho^Q_{z\vphantom{'}'}\big]^k$, $k \in \mathbbm{N}$ & Convex & Eq.~\eqref{eq:FPLower} & Eq.~\eqref{eq:FPUpper}
			\end{tabular}
		\end{ruledtabular}
		\caption{Examples of ensemble-averaged quantities for which we derive upper and lower bounds, and references to the relevant inequalities derived below. Here, $Q_1$ denotes a subsystem of $Q$, $\mathcal{N}$ is an arbitrary linear map over the space of operators, and the notation $\mathcal{N} \succeq 0$ indicates that this map is semi-positive definite, i.e.~$\llangle X|\mathcal{N}|X\rrangle \geq 0$ for all operators $X$. In some cases, two bounds are quoted, one of which is simple to evaluate, with the other being more versatile and amenable to optimization. The asterisk indicates that the bound is vacuous.}
		\label{tab:Bounds}
	\end{table*}
	\egroup
	
	Having outlined the general structure of our optimization-based approach in the previous section, we can now consider a range of different averaged properties that are of particular physical relevance, and construct explicit upper and lower bounds for each. Because the minimization and maximization problems have distinct characters, it is helpful to consider the two separately for each observable. The collection of quantities considered here is by no means exhaustive, and analogous bounds can be constructed for other observables, but the examples we choose encompass a broad range of quantities that are pertinent to measurement-induced dynamics.
	
	The logical arguments used to derive bounds for each quantity follow much the same pattern, and so after deriving the first several cases, we will simply quote the remainder of our results, leaving the proofs to Appendix \ref{app:Bounds}. Readers who are mainly concerned with the results of our calculations, rather than the detailed derivations, may skip the bulk of this section, and instead consult Table \ref{tab:Bounds}, where references to specific bounds are listed.
	
	\subsection{Global purity lower bound \label{subsec:PurityLB}}
	
	One of the simplest possible nonlinear averages that we can consider is the averaged global purity, $G(\rho^Q_z) = \Tr[(\rho^Q_z)^2]$ (to be distinguished from the purity of a subsystem of $Q$, which we treat in Section \ref{subsec:quadL}). As mentioned above, this is a convex function of $\rho^Q_z$, and hence the minimization problem can be solved using convex optimization techniques.
	The dual function \eqref{eq:DualFunctionMin} for this problem can be formally defined as
	\begin{align}
		h_-(\lambda_i) &=  \inf_{\blockstyle{\rho} \in \mathcal{K}_0} \sum_z p_z\Bigg(\Tr[(\rho^Q_z)^2] - \sum_i\lambda_i \Tr[\rho^Q_z A^{(i)}_z]\Bigg) \nonumber\\ &+ \sum_i \lambda_i b_i.
	\end{align}
	By virtue of the product structure of $\mathcal{K}_0 \equiv \mathcal{D}^{\oplus |\mathcal{Z}|}$, we can minimize with respect to each conditional state $\rho_z^Q$ separately, giving
	\begin{align}
		\label{eq:PurityMinF2}
		h_-(\lambda_i) &= \sum_z p_z F_{2,-}\left({\textstyle \sum_i} \lambda_i \tilde{A}_z^{(i)}\right) + \sum_i \lambda_i(b_i - a_i)
	\end{align}
	where $a_i \coloneqq \sum_z p_z \Tr[A_z^{(i)}]$, and $\tilde{A}_z^{(i)} = A_z^{(i)} - \Tr[A_z^{(i)}]\mathbbm{I}/d$ is the traceless part of $A_z^{(i)}$. Here we have defined a function over the space of traceless Hermitian operators $C$
	\begin{align}
		F_{2,-}(C) \coloneqq \inf_{\rho \in \mathcal{D}}\Big( \Tr[\rho^2] - \Tr[\rho C]\,\Big).
		\label{eq:OptimF2Purity}
	\end{align}
	Evidently, the above depends only on the eigenvalues of $C$. In principle, a fully general expression for the above can be found, however to make progress in the following we will use a simple lower bound $F_{2,-}(C) \geq 1/d - \Tr[C^2]/4$. (This actually becomes an equality if $|\min \text{eig} C| \leq 2/d$.)
	Invoking this bound, we are left with a manageable expression for the dual function
	\begin{align}
		h_-(\lambda_i) \geq -\frac{1}{4}\sum_z p_z \Tr\Big[\big({\textstyle \sum_i} \lambda_i \tilde{A}^{(i)}_z\big)^2\Big] + \sum_i \lambda_i(b_i-a_i)
		\label{eq:PurityMinLambda}
	\end{align}
	The right hand side is readily maximized over the Lagrange parameters $\lambda_i$, which gives us a lower bound for the solution of the dual problem
	\begin{align}
		h_-^* \geq \tilde{h}_- \coloneqq \frac{1}{d} + \sum_{ij}(b_i-a_i) [L^{-1}]_{ij}(b_j-a_j),
		\label{eq:PurityMinimumL}
	\end{align}
	where we define the matrix
	\begin{align}
		L_{ij} = \sum_z p_z \Tr[\tilde{A}_z^{(i)} \tilde{A}_z^{(j)}]
		\label{eq:PurityLij}
	\end{align}
	The lower bound \eqref{eq:PurityMinimumL} corresponds to the set of dual parameters $\lambda_i = 2 \sum_j [L^{-1}]_{ij}b_j$, and thus the bound becomes optimal ($\tilde{h}_- = h_-^*$) if the condition $\min \text{eig}(\sum_{ij} [L^{-1}]_{ij}b_j A^{(i)}_z) \geq -1/d$ is met for all $z$. This condition may or may not be met for any given set of measurements; regardless, Eq.~\eqref{eq:PurityMinimumL} constitutes a viable certificate by way of strong duality \eqref{eq:CertificateIneqs}, in that the averaged global purity of the true ensemble $\mathcal{E}^Q$ can be no less than the right hand side, i.e.~$\mathbbm{E}_z \Tr[(\rho^Q_z)^2] \geq \tilde{h}_-$.
	
	\subsubsection*{Incorporating classical simulations}
	
	Having reached this point, we can revisit the original problem and ask: how could we chosen $A_z^{(i)}$ in the first place in order to make our bound \eqref{eq:PurityMinimumL} as close as possible to the true value of the average $\mathbbm{E}_zG(\rho^Q_z)$? We can use the discussion of Section \ref{sec:Simulations} to guide our intuition. There, we posited that the quantum-classical correlator \eqref{eq:CorrQC}, which is a measurable quantity, would serve as a good proxy for the quantum-quantum correlator \eqref{eq:CorrQQ}, on account of the fact that the two coincide in the limit of perfect classical simulation $\rho^Q_z = \rho^C_z$. Here, we can define an analogous `quantum-classical purity',
	\begin{align}
		P^{QC} \coloneqq \sum_z p_z \Tr[\rho^Q_z \rho^C_z],
		\label{eq:PQC}
	\end{align}
	which can evidently be cast in the form of an estimable quantity \eqref{eq:LinearObservableZ}, and indeed is equal to the desired average when $\rho^Q_z = \rho^C_z$. Accordingly, it is instructive to consider an example where this is the only constraint we use, i.e.~$R = 1$, with $A^{(1)}_z = \rho^C_z$ and $b_1 = P^{QC}$. Then, altogether, the bound \eqref{eq:DualMaximumParallel} reduces to
	\begin{align}
		\mathbbm{E}_z\Big[\Tr\big[(\rho^Q_z)^2\big]\Big] \geq \frac{(P^{QC})^2}{P^{CC}},
		\label{eq:IneqCauchySchwartzPurity}
	\end{align}
	where we define $P^{CC} = \sum_z p_z \Tr[(\rho^C_z)^2]$ by analogy to Eq.~\eqref{eq:PQC}. The above inequality can also be proved by independent means using the Cauchy-Schwartz inequality, first applied to the Hilbert-Schmidt inner product $\Tr[\rho^Q_z \rho^C_z] \leq \sqrt{\Tr[(\rho^Q_z)^2]\Tr[(\rho^C_z)^2]}$, and then to the average $(\mathbbm{E}_z \sqrt{a_z b_z})^2 \leq (\mathbbm{E}_z a_z)(\mathbbm{E}_z b_z)$. This serves as a useful sanity check for our series of bounds $\tilde{h}_- \leq g_-^* \leq \mathbbm{E}_z \Tr[(\rho^Q_z)^2]$. We observe that as the quality of the classical simulation improves, the corresponding lower bound should increase, resulting in tighter bounds.
	
	We emphasise, however, that Eq.~\eqref{eq:PurityMinimumL} is far more versatile as a bound than the simple inequality \eqref{eq:IneqCauchySchwartzPurity}. In particular, we can incorporate multiple constraints \eqref{eq:Consistent} which allows us to use more information than just the averaged overlap between classical and quantum states. In particular, since the purity is a quadratic function of the density matrix, this suggests using more general quantum classical correlators that are bilinear in $\rho^Q_z$ and $\rho^C_z$. For this purpose, we introduce the superoperators (linear maps over the space of operators)
	\begin{subequations}
		\begin{align}
			\eta^{QC} &= \sum_z p_z |\rho^Q_z\rrangle \llangle \rho^C_z| \\
			\eta^{CC} &= \sum_z p_z |\rho^C_z\rrangle \llangle \rho^C_z|
		\end{align}
		\label{eq:CQSuperoperators}
	\end{subequations}
	Thinking of these as $(d^2 \times d^2)$-dimensional matrices, one can see that one can extract all possible quantum-classical and classical-classical correlators from the above objects: For any operators $A$, $B$, one has $\llangle A|\eta^{QC}|B\rrangle = \braket{A^\dagger \otimes B}^{QC}$, and similar for $\eta^{CC}$. Therefore, if we were to construct a complete basis of operators $\{\sigma^\mu\}$ and measure all correlators of the form $\braket{\sigma^\mu \otimes \sigma^\nu}^{QC}$ for $\mu, \nu = 1, \ldots, d^2$, we could fully reconstruct $\eta^{QC}$. For each one of these measured correlators, we will have a Lagrange multiplier $\lambda_i$, and these can also be organized into a superoperator form which we call $\tilde{\zeta}^{CQ}$. When arranged in this way, the dual function \eqref{eq:PurityMinLambda} then becomes
	\begin{align}
		h_-(\tilde{\zeta}^{CQ}) = 2\,\text{STr}[\eta^{QC}\tilde{\zeta}^{CQ}] - \text{STr}[(\tilde{\zeta}^{CQ})^\dagger \tilde{\zeta}^{CQ}].
		\label{eq:DualFnCQSuperoperator}
	\end{align}
	Here, $\text{STr}[\eta]$ denotes the trace of a superoperator $\eta$, which we could write in terms of a complete orthonormal basis of operators $\sigma_\mu$ as $\text{STr}[\eta] = \sum_\mu \llangle \sigma_\mu|\eta| \sigma_\mu\rrangle$. While the above holds for any choice of $\tilde{\zeta}^{CQ}$, by analogy to Eq.~\eqref{eq:PurityMinimumL} we can find the optimum choice $\zeta^{CQ}$ (written without a tilde), which is the solution to the superoperator equation (which is guaranteed to exist)
	\begin{align}
		\eta^{CC}\zeta^{CQ} = (\eta^{QC})^\dagger.
		\label{eq:ZetaQCImplicit}
	\end{align}
	(If $\eta^{CC}$ has an inverse, we could write $\zeta^{CQ} = (\eta^{CC})^{-1}(\eta^{QC})^\dagger$.) At this dual-optimum point, we obtain the bound
	\begin{align}
		\mathbbm{E}_z \Tr[(\rho^Q_z)^2] \geq g_-^* \geq \text{STr}\big[ \eta^{QC} \zeta^{CQ}\big]. 
		\label{eq:PurityLBSuperoperator}
	\end{align}
	Eq.~\eqref{eq:PurityLBSuperoperator} is our first concrete inequality that can be straightforwardly calculated using classical-quantum correlators. Evidently, in the case of perfect classical simulation $\rho^C_z = \rho^Q_z$, we have $\zeta^{CQ} = \text{id}$, and the inequalities in \eqref{eq:PurityLBSuperoperator} are both saturated. Hence, although we have lost some tightness in our calculations, we expect the bound to be near-optimal when the classical simulation is not quite perfect. We will benchmark how tight these inequalities are in various scenarios in Section \ref{sec:Benchmarking}.
	
	\subsection{Global purity upper bound \label{subsec:PurityUB}}
	
	In searching for an upper bound for the global purity, we could in principle set up the maximization problem \eqref{eq:Optimize} in an entirely analogous way to the above, finding an upper bound for $h_+^*$ and exploiting weak duality to obtain a corresponding bound for $g_+^*$. However, the global purity has a special significance in this context, which means that an immediate answer can be obtained by invoking the corollary of Theorem \ref{thm:ExtremePoints}: We know that there exists at least one ensemble consistent with the measurements \eqref{eq:Consistent} for which no more than $R$ of the states $\rho_z^Q$ are mixed. The existence of such an ensemble immediately implies that $g_+^*$ must be at least
	\begin{align}
		g_+^* \geq 1 - \left(1 - \frac{1}{d}\right)R \max_z p_z.
		\label{eq:PurityUB}
	\end{align}
	In the no-coincidence limit $p_z \ll 1$, the above very close to 1, which is itself a universally applicable upper bound for the averaged purity. We conclude that regardless of which observables $A_z^{(i)}$ we measure, we cannot obtain a non-vacuous upper bound for the averaged global purity. This fundamental obstruction is related to the special significance that global purity has for the structure of $\mathcal{K}$, and in particular its extreme points. We discuss this issue in depth in Section \ref{subsec:MixedCons}.

	\subsection{Lower bound for subsystem purities and quadratic observables \label{subsec:quadL}}
	
	Moving beyond the global purity, we can consider more general functions $G(\rho_z^Q)$ that are quadratic in the conditional states. Most generally, one can write these as
	\begin{align}
		G(\rho^Q_z) = \llangle \rho^Q_z| \mathcal{N}|\rho^Q_z\rrangle
		\label{eq:SqGSuperoperator}
	\end{align}
	where $\mathcal{N}$ is a Hermitian superoperator, i.e.~a linear map over the space of operators satisfying $\llangle C|\mathcal{N}|D\rrangle = \llangle D|\mathcal{N}|C\rrangle^*$ for operators $C, D$. To ensure the convexity of $G$, we will insist that $\mathcal{N}$ is positive semi-definite: $\llangle C|\mathcal{N}|C\rrangle \geq 0$ for any operator $C$; we write this condition as $\mathcal{N} \succeq 0$. Examples of this include the quantum-quantum correlator $\braket{O\otimes O}^{QQ}$ [Eq.~\eqref{eq:CorrQQ}], which corresponds to $\mathcal{N} = |O\rrangle \llangle O|$. Moreover, the purity of some subsystem of $Q$, which we will denote $Q_1$ with complement $Q_2 = Q \backslash Q_1$, can be written in this form: One can express purity of $Q_1$ as
	\begin{align}
		\Tr_{Q_1}[(\Tr_{Q_2}[\rho^Q_z])^2] = \sum_{\nu \in Q_1}  \llangle \rho^Q_z|\sigma_\nu\rrangle \llangle \sigma_\nu|\rho^Q_z\rrangle
		\label{eq:SubsystemPurityExpand}
	\end{align}
	where $\{\sigma_\nu\}$ is a basis of operators that respects the tensor product structure of the Hilbert space $\mathcal{H}^Q = \mathcal{H}^{Q_1} \otimes \mathcal{H}^{Q_2}$, and the notation $\nu \in Q_1$ denotes that the sum is restricted to those $\nu$ for which $\sigma_\nu$ acts as the identity on $Q_2$. The above implicitly defines the superoperator $\mathcal{N}$ corresponding to the subsystem purity.
	
	To solve the minimization problem for this class of observables, we use similar logic to that described in Section \ref{subsec:PurityLB}, with some modifications. Detailed arguments are presented in Appendix \ref{app:QuadLower}, which lead to the following bound
	\begin{align}
		\sum_z p_z \llangle \rho^Q_z| \mathcal{N}|\rho^Q_z\rrangle \geq \text{STr}\big[\mathcal{N}\eta^{QC} \zeta^{CQ}\big]
		\label{eq:QuadLower}
	\end{align}
	where $\zeta^{CQ}$ is the superoperator that was defined in Eq.~\eqref{eq:ZetaQCImplicit}. The global purity bound \eqref{eq:PurityLBSuperoperator} corresponds to the special case where $\mathcal{N} = \text{id}$, the identity superoperator. Again, the above inequality is saturated in the limit of perfect classical simulation.
	
	\subsection{Upper bound for quadratic observables \label{subsec:quadU}}
	
	The corresponding maximization problem for averages of the form \eqref{eq:SqGSuperoperator} is not a convex optimization problem, and so cannot be solved in a fully analogous way. Rather than directly solving the dual problem, we instead choose to re-express the problem by first trivially rewriting
	\begin{align}
		\label{eq:NRewrite}
		\mathcal{N} = \|\mathcal{N}\|_\infty \text{id} + \overline{\mathcal{N}},
	\end{align}
	where we define
	\begin{align}
		\overline{\mathcal{N}} = (\mathcal{N} - \|\mathcal{N}\|_\infty \text{id}),
	\end{align}
	and $\|\mathcal{N}\|_\infty = \max_{\llangle C|C\rrangle=1} \llangle C|\mathcal{N}|C\rrangle$ is the spectral norm of $\mathcal{N}$ when viewed as a matrix. The significance of the above is that $\overline{\mathcal{N}} \preceq 0$ by construction, and hence when we substitute this into Eq.~\eqref{eq:SqGSuperoperator}, the second term constitutes a concave function of $\rho^Q_z$. The first term simply gives us a term proportional to the purity $\Tr[(\rho^Q_z)^2]$, which by Theorem \ref{thm:ExtremePoints} we know to be very close to unity at the point where the maximum is achieved. Hence, we lose little tightness by replacing the first term with the constant $\|\mathcal{N}\|_\infty$.
	
	Being concave, the average of $\llangle \rho^Q_z | \overline{\mathcal{N}}|\rho^Q_z\rrangle$ can be upper bounded, using the same method as the lower bound for the convex function \eqref{eq:QuadLower}. Altogether, we obtain
	\begin{align}
		\mathbbm{E}_z \llangle \rho^Q_z|\mathcal{N}|\rho^Q_z\rrangle &\leq \|\mathcal{N}\|_\infty + \text{STr}[\overline{\mathcal{N}}\eta^{QC} \zeta^{CQ}] \nonumber\\
		&=\|\mathcal{N}\|_\infty\big(1 - \text{STr}[\eta^{QC} \zeta^{CQ}]\big) \nonumber\\ &+\text{STr}[\mathcal{N} \eta^{QC} \zeta^{CQ}].
		\label{eq:QuadUpper}
	\end{align}
	From Eqs.~(\ref{eq:QuadLower}, \ref{eq:QuadUpper}), it becomes clear that we can constrain the average of \eqref{eq:SqGSuperoperator} to within a window whose width is determined by the quantity $(1 - \text{STr}[\eta^{QC} \zeta^{CQ}])$. Indeed, from Sections \ref{subsec:PurityLB} and \ref{subsec:PurityUB}, this quantity is itself equal to the range of values that the global purity can take. This highlights the significance of Theorem \ref{thm:ExtremePoints}: If the true states $\rho^Q_z$ realised by the device are themselves mixed, then the purity cannot be constrained to be within an arbitrarily narrow window, and in turn, quadratic observables of this kind cannot be tightly constrained either. Indeed, even before we performed any of the manipulations in this section, one could straightforwardly show that
	\begin{align}
		\text{ran}(\mathcal{N}) + \text{ran}(\overline{\mathcal{N}}) &\geq \|\mathcal{N}\|_\infty \text{ran}(\text{id}) \nonumber\\ &= \|\mathcal{N}\|_\infty\big(1 - \text{STr}[\eta^{QC} \zeta^{CQ}]\big)
	\end{align}
	where we use the shorthand $\text{ran}(\mathcal{N}) = g_+^* - g_-^*$ for the observable $\mathbbm{E}_z[\llangle \rho^Q_z |\mathcal{N}|\rho^Q_z\rrangle]$. Hence, if there is large uncertainty in the purity, then there must also be large uncertainty in certain quadratic observables (see also Section \ref{subsec:MixedCons}).

	\subsection{von Neumann entropy upper bound}
	
	The von Neumann entropy $S(\rho^Q_z) = -\Tr[\rho^Q_z \log \rho^Q_z]$ is another quantity that can be used to characterized mixedness and/or entanglement of quantum states, which has particular information-theoretic significance. This function is concave in $\rho^Q_z$, and so in this instance the maximization problem is most easily addressed. 
	The dual function for the maximization problem is
	\begin{align}
		h_+(\lambda_i) = \sum_z p_z F_{S, +}\Big({\textstyle \sum_i} \lambda_i A_z^{(i)}\Big) + \sum_i \lambda_ib_i ,
	\end{align}
	where by analogy to $F_{2,-}(C)$, we define $F_{S, +}(C) = \sup_{\rho^{Q} \in \mathcal{D}} (S(\rho^{Q}) - \Tr[\rho^{Q} C])$. A fairly straightforward calculation reveals
	\begin{align}
		F_{S, +}(C) = \log \Tr[e^{-C}],
		\label{eq:vNMaxSubProblem}
	\end{align}
	which is attained for $\rho = e^{-C}/\Tr[e^{-C}]$. Hence, we have
	\begin{align}
		\mathbbm{E}_z S(\rho^{Q}_z) \leq 
		h_+^* &= \min_{\lambda_i} \bigg(\sum_i \lambda_i b_i \nonumber\\
		&+\sum_z p_z \log \Tr_{Q}\Big[\exp\big(-{\textstyle \sum_i}\lambda_iA_z^{(i)}\big)\Big]\bigg)
		\label{eq:vNMaxLambda}
	\end{align}
	This upper bound can be applied quite generally, and can even be used to re-derive a result that was obtained in Ref.~\cite{Garratt2023a} using alternative arguments based on the quantum relative entropy: If we have a single constraint with $A_z = -\log \rho^{C}_z$, then
	\begin{align}
		h_+(\lambda) = \sum_z p_z \Big(\log \Tr[(\rho^{C}_z)^\lambda] - \lambda \Tr[\rho^{Q}_z \log \rho^{C}_z]\Big).
		\label{eq:vNMaxDualFunction}
	\end{align}
	Setting $\lambda = 1$ in the above, we obtain a certificate
	\begin{align}
		\mathbbm{E}_z S(\rho^{Q}_z) \leq h_+(\lambda=1) = -\sum_z p_z \Tr[\rho^{Q}_z \log \rho^{C}_z],
		\label{eq:vNMaxQC}
	\end{align}
	the right hand side of which was introduced in Ref.~\cite{Garratt2023a} as the `quantum-classical entanglement entropy'. Evidently, we could use exactly the same inequality for the average von Neumann entropy of a subsystem of $Q$, provided one replaced all density matrices with the corresponding reduced density matrices. The significance of the choice $\lambda = 1$ can be understood by recognising that $\lambda = 1$ is dual-optimal in the limit of perfect simulation $\rho^C_z = \rho^Q_z$, since the inequality \eqref{eq:vNMaxQC} is then saturated. If there are small discrepancies in the quantum and classical states, one expects that the optimal choice of $\lambda$ will be shifted slightly, and hence this bound can in principle be tightened by optimizing over $\lambda$.
	
	However, a more serious problem to address is the instability of the quantum-classical entanglement entropy for singular or near-singular classical states $\rho^C_z$, since $\log \rho^C_z$ diverges when the eigenvalues of $\rho^C_z$ are small. This is a particularly important problem when $\rho^C_z$ are close to being pure, as noted in Ref.~\cite{Garratt2023a}. Here, our more general approach allows one to get around this issue in a systematic way, since the choice $A_z = -\log \rho^C_z$ can be easily altered in a way that guarantees numerical stability. A natural choice of regularization is to work directly with the eigenvalue decomposition $\rho^C_z = \sum_n q_{z,n}\ket{\chi_{z,n}}\bra{\chi_{z,n}}$, and to separate out the problematic eigenvalues---namely those that are below some threshold $\epsilon$. We are free to separately measure the following two quantities
	\begin{align}
		A_z^{(1)} &= -\Pi^>_z \log \rho^C_z \Pi^>_z, & A_z^{(2)} = \Pi_z^{\leq},
		\label{eq:vNRegularizedObservables}
	\end{align}
	where we define the projector $\Pi^\leq_z = \sum_{q_n \leq \epsilon} \ket{\chi_{z,n}}\bra{\chi_{z,n}}$ and its complement $\Pi^{>}_z = \mathbbm{I}-\Pi^\leq_z$. The first of these is similar to the original choice $A_z = -\log \rho^C_z$ but with near-singular eigenvalues removes, while the second measures the average weight of $\rho^Q_z$ lying within the near-singular subspace $\Pi_z^{\leq}$. We can then proceed as before and find an optimal bound based on empirical values of the above two observables. In Appendix \ref{app:vNUpper}, we work out such an upper bound explicitly. As a particularly simple special case, if the classical states are all pure $\rho^C_z = \ket{\phi^C_z}\bra{\phi^C_z}$, then this improved bound can be expressed in terms of the quantity $\overline{\delta^Q} = 1 - \sum_z p_z \braket{\phi^C_z|\rho^Q_z|\phi^C_z}$ as
	\begin{align}
		\mathbbm{E}_z S_{\rm vN}(\rho^Q_z) \leq H_2(\overline{\delta^Q}) + \overline{\delta^Q}\log(d-1),
		\label{eq:vNUpperPure}
	\end{align}
	where $H_2(p) = -p\log p - (1-p)\log(1-p)$ is the Shannon entropy for a binary random variable.

	\subsection{von Neumann entropy lower bound}
	
	If we focus on the global von Neumann entropy, then arguments similar to those in Section \ref{subsec:PurityUB} can be used to show that a non-vacuous lower bound cannot be obtained: Theorem \ref{thm:ExtremePoints} implies the existence of a feasible ensemble whose average entropy is no greater than $R \max_z p_z \log d$, which is small in the no-coincidence limit $p_z \ll 1$.
	
	As for the von Neumann entropy of a subsystem $Q_1$, this can in principle be lower bounded, but a direct analysis of the minimization problem is not straightforward, due to the concavity of $S(\rho^{Q_1}_z)$. Instead we use a similar approach to Subsection \ref{subsec:quadU}, where we exploited the fact that the maximum of a convex function is attained at an extreme point, which we know to be a (mostly) pure ensemble. There, we rewrote the desired function in terms of the difference between the purity and a convex function [Eq.~\eqref{eq:NRewrite}], thereby allowing the maximization problem transformed into a convex optimization task. Here, we use a similar line of reasoning, by trivially rewriting
	\begin{align}
		S(\rho^{Q_1}_z) = S(\rho^{Q}_z) - S(\rho^Q_z|Q_1)
		\label{eq:CondEntropyDef}
	\end{align}
	where $S(\rho^{AB}|B) \coloneqq S(\rho^{AB}) - S(\rho^B)$ is the conditional quantum entropy for a bipartite state $\rho^{AB}$. From Theorem \ref{thm:ExtremePoints} and the concavity of $S(\rho^{Q_1}_z)$, the desired minimum is attained for an ensemble that is mostly globally pure, and so the average of $S(\rho^Q_z)$ will be close to 0 at this point. This suggests that we do not lose much tightness in employing the bound
	\begin{align}
		g_-^* = \min_{\blockstyle{\rho}\in \mathcal{K}} \mathbbm{E}_z S(\rho^{Q_1}_z) &\leq \min_{\blockstyle{\rho}\in \mathcal{K}}\Big[\!-\mathbbm{E}_zS(\rho^Q_z|Q_1)\Big] \nonumber\\ &= -\max_{\blockstyle{\rho}\in \mathcal{K}}\Big[\mathbbm{E}_zS(\rho^Q_z|Q_1)\Big]
		\label{eq:vNConditionalIneqs}
	\end{align}
	Crucially, $S(\rho^Q_z|Q_1)$ is a concave function of the global state $\rho^Q_z$ (a statement that is equivalent to strong subadditivity of the von Neumann entropy \cite{Nielsen2010}). Thus, we are left with a convex optimization problem, as desired. In Appendix \ref{app:vNLower}, we derive the following concrete bound
	\begin{align}
		\mathbbm{E}_z\big[S(\rho_z^{Q_1})\big] &\geq { \sum_i}\lambda_i b_i - \mathbbm{E}_z \log \|\Tr_{Q_2} e^{\sum_i \lambda_i A_z^{(i)}} \|_\infty
		\label{eq:vNMinInfNorm}
	\end{align}
	for any choice of the Lagrange multipliers $\lambda_i$. Again, we can specialise to a particular choice of operators $A_z^{(i)}$ to re-derive a result of Ref.~\cite{Garratt2023a}: Taking $A^{(1)}_z = -\log \rho^C_z$ and $A^{(2)} = +\log \rho^{C_1}_z$, one can set $\lambda_1 = \lambda_2 = 1$. With this particular choice, the argument of the logarithm in the above becomes $\|\Tr_{Q_2} e^{\log \rho^C_z - \log \rho^{C_1}_z\otimes \mathbbm{I}^{Q_2}}\|_\infty$, which can be shown to be no greater than unity using Theorem 11.29 of Ref.~\cite{Petz2008}; we then have
	\begin{align}
		\mathbbm{E}_z\big[S(\rho_z^{Q_1})\big] &\geq \mathbbm{E}_z\Big[\Tr[\rho^Q_z\log \rho^C_z] - \Tr[\rho^{Q_1}\log \rho^{C_1}_z]\Big]
		\label{eq:vNMinSimple}
	\end{align}
	As with the other optimization problems considered in this section, we can choose to use bounds that are simple and easy to evaluate, such as Eq.~\eqref{eq:vNMinSimple}, or to use the more versatile expression \eqref{eq:vNMinInfNorm}, which are more cumbersome, but can in principle be numerically optimized to obtain a tighter inequality.
	
	\subsection{Frame potential}
	
	The frame potential is a property of a quantum state ensemble that characterises the variation between different quantum states in the ensemble. For any integer $k$, it is defined as
	\begin{align}
		F^{(k)}(\mathcal{E}^Q) \coloneqq \sum_{z,z'}p_zp_{z'}\Tr[\rho^Q_z \rho^Q_{z'}]^k
		\label{eq:FPDef}
	\end{align}
	For pure state ensembles, the frame potential can be used to characterize how far away an ensemble is from being a quantum state $k$-design \cite{Roberts2017,Ippoliti2022}, namely an ensemble for which the $k$th moments $\sum_z p_z (\rho^Q_z)^{\otimes k}$ coincide with those of the Haar ensemble \cite{Renes2004,Ambainis2007} (see Section \ref{subsec:designs}).
	
	Although the quantity \eqref{eq:FPDef} is not manifestly in the form of the averages \eqref{eq:AverageProperty} considered so far, we can use a simple trick to bring it into the appropriate form: Focusing on $k = 2$ for now, we can write $F^{(2)} = \sum_z p_z \llangle \rho^Q_z|\eta^{QQ}|\rho^Q_z\rrangle$, where $\eta^{QQ}$ is defined by analogy to Eq.~\eqref{eq:CQSuperoperators}. We do not have access to $\eta^{QQ}$ in advance, but from the bounds (\ref{eq:QuadLower}, \ref{eq:QuadUpper}) we can infer that
	\begin{subequations}
		\begin{align}
			\eta^{QQ} &\succeq \frac{\eta^{QC}\zeta^{CQ} + \text{H.c.}}{2}  \\
			\eta^{QQ} &\preceq \frac{\eta^{QC}\zeta^{CQ} + \text{H.c.}}{2}  + (1 - \text{STr}[\eta^{QC} \zeta^{CQ}])\times \text{id}
		\end{align}
	\end{subequations}
	where the notation $\mathcal{N} \preceq \mathcal{N}'$ for two superoperators $\mathcal{N}$, $\mathcal{N}'$ means $\llangle C|\mathcal{N}|C\rrangle \leq \llangle C|\mathcal{N}'|C\rrangle$ for all operators $C$. Moreover, by its definition we have $\|\eta^{QQ}\|_\infty = \max_{\|C\|_2^2 = 1} \sum_z p_z \Tr[\rho^Q_z C]^2 \leq 1$. Hence, successive application of the bounds (\ref{eq:QuadLower}, \ref{eq:QuadUpper}) yield
	\begin{subequations}
		\begin{align}
			F^{(2)}(\mathcal{E}^Q) &\geq \text{STr}[(\eta^{QC}\zeta^{CQ})^2] \label{eq:FPLower} \\
			F^{(2)}(\mathcal{E}^Q) &\leq \text{STr}[(\eta^{QC}\zeta^{CQ})^2]
			+ (1 - \text{STr}[\eta^{QC}\zeta^{CQ}]^2).
			\label{eq:FPUpper}
		\end{align}
		\label{eq:FPBounds}
	\end{subequations}
	
	More generally, for higher $k$ the frame potential can be expressed in terms of a new `doubled' ensemble $\mathcal{E}^{QQ}$ with the following structure
	\begin{align}
		\mathcal{E}^{QQ} = \{(p_zp_{z'}, \rho^Q_z \otimes \rho^Q_{z'})\}_{(z,z') \in \mathcal{Z}\times \mathcal{Z}}
	\end{align}
	In words, samples from this ensemble correspond to independently sampled pairs $(z,z')$ of the original ensemble $\mathcal{E}^Q$, and the corresponding states are tensor products $\rho^Q_z \otimes \rho^Q_{z'}$. Then, the frame potential \eqref{eq:FPDef} of $\mathcal{E}^Q$ can be interpreted as an average of the form \eqref{eq:AverageProperty} for $\mathcal{E}^{QQ}$, where the function $G$ is given by
	\begin{align}
		G(\sigma) = \Tr[\sigma \pi_S]^k
	\end{align}
	where $\sigma$ is a density operator on the doubled space, and $\pi_S = \sum_{a,b}\ket{a\otimes b}\bra{b \otimes a}$ is the swap operator. Bounds for averages of $k$th powers of expectations, such as the above, can in principle be derived using similar approaches to those described in this Section.
	
	\section{Numerical benchmarking\label{sec:Benchmarking}}
	
	Having derived various bounds for certain ensemble-averaged quantities of interest, we now present results of some numerical experiments where we simulate the full procedure described in this work, including estimation of the measurable parameters \eqref{eq:Consistent} to constructing the bounds.
	
	\subsection{Projected ensemble}
	
	As a testbench for our method, we will consider the projected ensemble of a many-body state generated by (possibly noisy) finite time evolution from a product state. For the noiseless case, we take as the pre-measurement state $\ket{\Psi(t)} = U(t)\ket{0^{\otimes N}}$, where $t$ is an integer-value discrete time, and $U(t) = U_F^t$ is a unitary generated by Floquet evolution with Floquet unitary $U_F = e^{-\iu t_2 H_2} e^{-\iu t_1 H_1}$. We choose the Hamiltonians $H_{1,2}$ to be given by the one-dimensional tilted-field Ising model with open boundary conditions $H_{\alpha} = \sum_{j=1}^{N-1} J_{j,\alpha} Z_jZ_{j+1} + h^x_{j,\alpha}X_j + h^z_{j,\alpha} Z_j$ for $\alpha = 1, 2$, where $(X_j, Y_j, Z_j)$ are Pauli operators on site $j$. For the purposes of this subsection, we fix $J_{j,\alpha} = 1$, $h^z_{j,\alpha} = (1+\sqrt{5})/2$, $h^x_{j,1} = 0.4$, $h^x_{j,2} = -0.6$. The projected ensemble considered here is defined as the set of post-measurement states arising when $N-1$ qubits are measured in the computational (Pauli-$Z$) basis, with the qubit on site $j = 1$ acting as the `unmeasured' qubit. In cases where noise is present, we will apply an amplitude damping channel of strength $p_{\rm dec}$ on every qubit after each Floquet period. Concretely, the system density matrix evolves as $\rho(t+1) = (\mathcal{N}_{\rm damp} \circ \mathcal{N}_{\rm unit})[\rho(t)]$, with $\mathcal{N}_{\rm unit}[\rho] = U_F \rho U_F^\dagger$, and $\mathcal{N}_{\rm damp}[\rho] = K_0\rho k_0^\dagger + K_1 \rho K_1^\dagger$, with $K_0 = \text{diag}(1, \sqrt{1-p_{\rm dec}})$ and $K_1 = \bigl(\begin{smallmatrix} 0 & \sqrt{p_{\rm dec}} \\ 0 & 0 \end{smallmatrix}\bigr)$. In the noisy case, where we must compute the full evolution of the density matrix, our exact diagonalization results are limited to relatively small system sizes. To ensure cross-comparability between all cases, we will fix $N = 10$ throughout. Note that the specific form of evolution we choose, along with the parameters selected, do not have a great deal of bearing on the performance on our method---indeed we have considered several different settings and found quantitatively similar performance. For the purposes of this section, we will focus on the particular average $\bar{G} = \mathbbm{E}_z \Tr[\rho^Q_z Z_1]^2$, for which we can use the bounds Eqs.~(\ref{eq:QuadLower}, \ref{eq:QuadUpper}); again, this choice is not particularly important.
	
	We will simulate the full protocol using the shadow tomography-based method described in Section \ref{subsec:Shadows} (Fig.~\ref{fig:shadow}) to extract the values of the estimable parameters \eqref{eq:LinearObservableZ}. Specifically, the procedure involves 1) Preparation of the pre-measurement state, 2) Applying random on-site Clifford unitaries to the unmeasured qubits, 3) Projectively measuring all qubits, 4) Processing the measurement outcomes via the dual frame \eqref{eq:DualFramePauli}, in terms of which estimators for the quantum-classical correlators can be obtained, and finally 5) Using the bounds derived in the previous section to constrain the value of the desired quantity. This is repeated a finite number of times $M$.
	
	The finite number of repetitions $M$ means that there will be some residual uncertainty in the values of the chosen estimable parameters. Taking the example of the correlators  \eqref{eq:CQSuperoperators}, we will obtain an estimate $\hat{\eta}^{QC} = \overline{\eta}^{QC} + \hat{\epsilon}^{QC}$, where $\overline{\eta}^{QC}$ is the true value, and $\hat{\epsilon}^{QC}$ is a zero-mean error, whose variance decays as $1/\sqrt{M}$ (hats are used to denote random variables here). Despite this uncertainty, we can still employ the bounds in a rigorous manner as follows: We use the estimated value $\hat{\eta}^{QC}$ as parameters for finding the optimal Lagrange multipliers [in this case we substitute $\hat{\eta}^{QC}$ into Eq.~\eqref{eq:ZetaQCImplicit} to find $\tilde{\zeta}^{CQ}$]. This will give us a set of Lagrange multipliers that are approximately dual-optimal point for the true problem, which features $\overline{\eta}^{QC}$ rather than $\hat{\eta}^{QC}$. These Lagrange multipliers can then be substituted into the dual function [in this case Eq.~\eqref{eq:DualFnCQSuperoperator}], which by virtue of being linear in $b_i$ (equivalently $\eta^{QC}$) can be evaluated with some error bars based on estimates of the standard deviation of $b_i$, along with some pre-chosen confidence level---we pick 99\% confidence throughout. One can then guarantee that the true value is no smaller or greater (as appropriate) than the constructed value with probability at least 99\%.\\
	
	\begin{figure}
		\centering
		\includegraphics{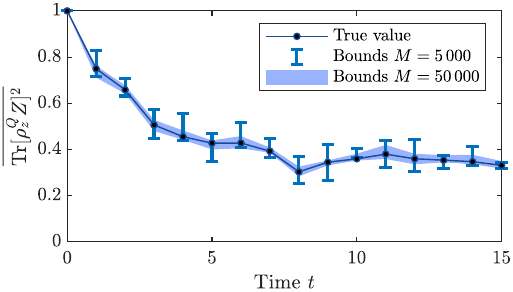}
		\caption{Estimates of the averaged quantity $\bar{G} = \mathbbm{E}_z(\Tr[\rho^Q_zZ_1]^2)$ for the projected ensemble generated from noiseless dynamics as described in the main text. The true value is shown, along with bounds constructed using our method for two different choices of the number of experimental repetitions $M = \num{5000}$ and $M = \num{50000}$. In this case, classical simulation is perfect and the states in the ensemble are pure, so as $M$ is increased both upper and lower bounds converge asymptotically towards the true value.}
		\label{fig:NoiselessPerfect}
	\end{figure}
	
	As a first check, we can consider the best-case scenario, where dynamics is noiseless and the classical simulation is perfect $\rho^C_z = \rho^Q_z$. We show upper and lower bounds for the target function $\bar{G} = \mathbbm{E}_z(\Tr[\rho^Q_zZ_1]^2)$ as calculated via our method, along with the true value, in Fig.~\ref{fig:NoiselessPerfect}. Due to the fact that the ensemble states are pure and simulation is perfect, the only source of uncertainty in the inferred value of $\bar{G}$ is from the finite number of samples $M$, which we deal with as described above. We show bounds empirically constructed for two different sample sizes $M$, and both lower and upper bounds will tend towards the true value as $M$ is increased indefinitely, with corrections scaling as $1/\sqrt{M}$.
	
	We can now introduce some inaccuracy in the classically simulated states $\rho^C_z$. For this purpose, we will consider the projected ensemble at a fixed time $t = 8$, using the same `quantum' states $\rho^Q_z$ as before. To construct the imperfect classical states, we will employ the same Floquet evolution as above, but now with some additional spatially-dependent randomness in each of the Hamiltonian parameters $J_{j,\alpha}$, $h_{j,\alpha}^{x,z}$. For each parameter, we pick a perturbed value $\tilde{J}_{j,\alpha} = J_{j,\alpha}(1 + f n_{j,\alpha})$ (similar for $\tilde{h}_{j,\alpha}^{x,z}$), where $n_{j,\alpha} = \pm 1/2$ are chosen independently at random for each parameter, but kept fixed in all of the data shown here. The free parameter $f$ represents a fractional uncertainty in the Hamiltonian parameters; thus as it is increased, the quantum and classical states $\rho^Q_z$, $\rho^C_z$ become less correlated. We can keep track of accuracy of the classical states in our numerical experiment by evaluating the quantity $\Delta^{QC} \coloneqq \sum_z p_z \|\rho^Q_z - \rho^C_z\|_1$, where $\|\rho - \sigma\|_1$ is the trace distance. 
	
	\begin{figure}
		\centering
		\includegraphics{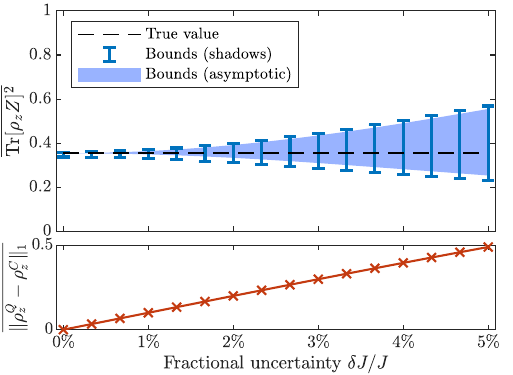}
		\caption{Effect of the fidelity of the classically simulated states $\rho^C_z$ on the tightness of the bounds for the noiseless projected ensemble at time $t = 8$. The classical states $\rho^C_z$ are perturbed away from the ensemble states $\rho^Q_z$ by introducing some fractional uncertainty $f \sim \delta J/J$ in the Hamiltonian parameters, as described in the main text. In the top panel, we include both bounds constructed based on shadow tomographic data from $M = \num{50000}$ repetitions, along with asymptotic bounds, where we evaluate the right hand sides of the inequalities (\ref{eq:QuadLower}, \ref{eq:QuadUpper}) without any statistical uncertainty---this is the value that one would obtain as $M \rightarrow \infty$. The bottom panel shows the average trace distance between quantum and classical states $\Delta^{QC} = \sum_z p_z \|\rho^Q_z - \rho^C_z\|_1$, which acts as a measure of how faithful the simulated states are.}
		\label{fig:innac_noiseless}
	\end{figure}
	
	Figure \ref{fig:innac_noiseless} shows data obtained using these inaccurate classical states. We show both the bounds constructed based on data from a finite number of experimental samples $M = \num{50000}$, along with `asymptotic' (i.e.~$M \rightarrow \infty$) bounds, which we obtain by evaluating the right hand sides of Eqs.~(\ref{eq:QuadLower}, \ref{eq:QuadUpper}), thus eliminating any statistical uncertainty. We see that as the fractional uncertainty $f$ is increased, the distance between the quantum and classical states increases, and the bounds become less tight. For reference, the value of $\Delta^{QC}$ that would be obtained if $\rho^C_z$ were chosen as independent random states would be 1; thus the data towards the right of this plot represents fairly poor simulation. In the absence of noise and for the specific quantity considered here, crude estimates for the deviation between the bounds and the true value scale as $\sum_z p_z\|\rho^Q_z - \rho^C_z\|^2$ (at least in the regime of small $\Delta^{QC}$), which is upper bounded by $(\Delta^{QC})^2$. This quadratic dependence on the distance between quantum and classical states explains why we see good performance, even when there is appreciable difference between the two states.
	
	Finally, we consider the case where both noise and inaccuracies in the classical states are present, which reflects the nature of realistic experiments. Again we fix $t = 8$, and now consider amplitude damping of strength $p_{\rm dec} = \num{0.002}$ for each qubit and each timestep. Although this value of $p_{\rm dec}$ appears to be small, the expected number of errors in the full circuit can be estimated to be $N\times t \times p_{\rm dec} = \num{0.24}$, which has an appreciable effect on the conditional states, as can be seen in their mean purity, which here is $\num{0.95}$. The classical states are generated in the same way as the ensemble states, including the noise, but with additional randomness in the Hamiltonian parameters of strength $f$, introduced in the same manner as above. 
	
	\begin{figure}
		\centering
		\includegraphics{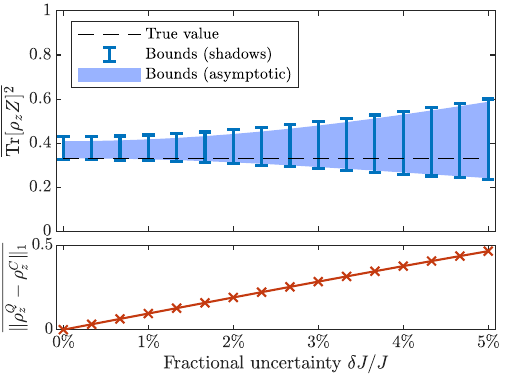}
		\caption{As in Figure \ref{fig:innac_noiseless}, but with noisy dynamics ($p_{\rm dec} = 0.002$), resulting in a projected ensemble with mixed states, with average purity $\sum_zp_z \Tr[(\rho^Q_z)^2] = \num{0.95}$. The lower bound approaches the true value in the limit of perfect simulation, but the upper bound does not, as a consequence of Theorem \ref{thm:ExtremePoints}.}
		\label{fig:noisy}
	\end{figure}
	
	The results are shown in Fig.~\ref{fig:noisy}. The data follows most of the same trends as before, with higher fractional uncertainty leading to higher deviation $\Delta^{QC}$, and in turn less tight bounds. However, because the dynamics is noisy, the states in the ensemble $\rho^Q_z$ are no longer pure---in this case their average purity is $\sum_zp_z \Tr[(\rho^Q_z)^2] = \num{0.95}$. Unlike the lower bound, the upper bound saturates at a value that is strictly above the true value of $\bar{G}$ even in the limit of perfect simulation ($f = 0$). This can be seen as a consequence of Theorem \ref{thm:ExtremePoints}: We saw already in Section \ref{subsec:PurityUB} that the global purity cannot be tightly bounded, and the same goes for other quadratic observables. This issue is discussed in detail in Section \ref{subsec:MixedCons}.

	\subsection{Application: Verifying emergent quantum state designs \label{subsec:designs}}
	
	One feature of the projected ensemble that has attracted much interest recently is that under rather generic conditions, the ensemble of post-measurement states turns out to form an (approximate) quantum state $k$-design \cite{Ho2022,Claeys2022,Cotler2023}. That is, the $k$th moments of the ensemble
	\begin{align}
		\rho^{(k)} \coloneqq \sum_z p_z (\rho^Q_z)^{\otimes k}
		\label{eq:EnsembleMomentK}
	\end{align}
	agree (closely) with the corresponding moments of the Haar ensemble
	\begin{align}
		\rho^{(k)}_{\rm Haar} \coloneqq \int \dif\mu_{\rm Haar}(\psi) \big(\ket{\psi}\bra{\psi}\big)^{\otimes k} = \frac{\Pi_{\rm sym}^{(k)}}{{k+d-1 \choose k}}
		\label{eq:EnsembleMomentHaar}
	\end{align}
	where $\Pi_{\rm sym}^{(k)}$ is the projector onto the permutation-symmetric subspace of $(\mathcal{H}^Q)^{\otimes k}$. Such ensembles of states are in a certain sense `maximally random', and this makes them useful for certain tasks including quantum state and channel tomography \cite{Hayden2004,Scott2006,Knill2008,Brandao2016,Mcginley2022a} and cryptography \cite{Radhakrishnan2009}. Here we demonstrate how our method can be employed to verifiably conclude whether or not the projected ensemble realised in an experiment is an approximate $k$-design.
	
	For ensembles of pure states, the frame potential constitutes a measure of how far the ensemble is from being a $k$-design. Specifically, in Ref.~\cite{Ippoliti2022} it was shown that the frame potential \eqref{eq:FPDef} is related to the (normalized) Frobenius distance between the moments (\ref{eq:EnsembleMomentK}, \ref{eq:EnsembleMomentHaar}) via
	\begin{align}
		\|\rho^{(k)} - \rho^{(k)}_{\rm Haar}\|_2^2 \;\stackrel{\mbox{\text{\footnotesize  (pure)}}}{=} \; F^{(k)}(\mathcal{E}^Q) - F^{(k)}_{\rm Haar}
		\label{eq:FPDistancePure}
	\end{align}
	with $F^{(k)}_{\rm Haar} = {k+d-1\choose k}^{-1}$ the frame potential for the Haar ensemble. Thus, when the frame potential is minimized, a $k$-design can be formed. However, in experiment the ensemble states can be mixed, and this leads to a suppression of the frame potential which can mimic the effect of forming a quantum state design---indeed, the right hand side of Eq.~\eqref{eq:FPDistancePure} can even be negative when the states are mixed. For this case, we need a more generally applicable bound. By expanding the left hand side of \eqref{eq:FPDistancePure}, we find that the distance can more generally be re-expressed as
	\begin{align}
		\|\rho^{(k)} - \rho^{(k)}_{\rm Haar}\|_2^2 = F^{(k)} + F^{(k)}_{\rm Haar}\left(1 - \frac{2}{k!}\sum_{\tau \in \Sigma_k} \mathbbm{E}_z[P_z^{(\tau)}]\right)
	\end{align}
	where we use the shorthand
	\begin{align}
		P_z^{(\tau)} = \prod_{c \in \tau} \Tr[(\rho^Q_z)^{|c|}]
	\end{align}
	Here, each $c \in \tau$ is a cycle of the permutation group element $\tau$, whose length is $|c|$. In the case $k = 2$, which we will focus on here, this gives
	\begin{align}
		\|\rho^{(2)} - \rho^{(2)}_{\rm Haar}\|_2^2 = F^{(k)}(\mathcal{E}^Q) - \frac{2}{d(d+1)}\mathbbm{E}_z\big[\!\Tr[(\rho^Q_z)^2]\big] 
		\label{eq:Dist2}
	\end{align}
	The right hand side of the above contains the frame potential and the average global purity. Interestingly, these appear with opposite signs, and thus we have a difference between two convex functions, which itself is not convex. Nevertheless, we can employ the upper and lower bounds  derived in Section \ref{sec:Constr} for each term separately. 
	
	\begin{figure*}
		\centering
		\includegraphics[scale=1]{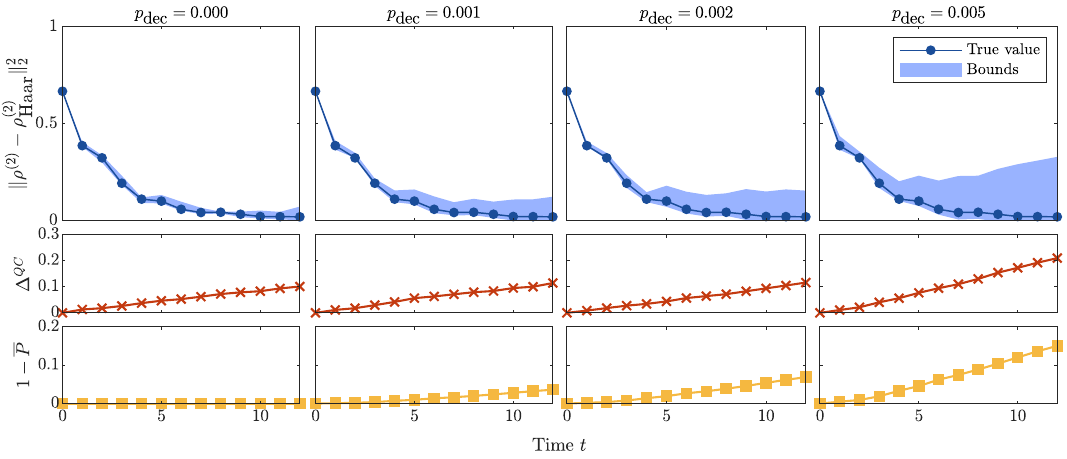}
		\caption{Distance \eqref{eq:Dist2} between the $k = 2$ moments of the projected ensemble from those of the Haar ensemble, as calculated using our method with $M = \num{50000}$ experimental repetitions (classical shadows), for various values of $p_{\rm dec}$ (top panel). Classical states are generated using a fractional parameter uncertainty $f = 0.5\%$, and without noise. For reference, we also show the average trace distance between quantum and classical states $\Delta^{QC}$ (middle panel), along with $1 - \sum_z p_z \Tr[(\rho^Q_z)^2]$, which measures how far the states are from being mixed on average (bottom panel).}
		\label{fig:design_dist}
	\end{figure*}
	
	We can test this method using the same dynamics that we considered in the previous subsection, including both noise and inaccuracies in the classical states. The tilted-field Ising Hamiltonian that generates the Floquet unitary is understood to be chaotic \cite{Kim2013}, and hence we expect to see emergent state designs in the projected ensemble \cite{Cotler2023}. The results of our numerical simulations are shown in Fig.~\ref{fig:design_dist}. We plot both the true value of the distance \eqref{eq:Dist2} along with bounds constructed from simulated experimental data using $M = \num{50000}$ repetitions. In this case, the classically simulated states are generated from noiseless evolution, but with a fractional uncertainty in the parameters of $f = 0.5\%$. Evidently, the distance does indeed decay towards zero as time increases. For small values of noise, the bounds we obtain are relatively tight, meaning that in a real experiment (where one would not have access to the true value), a definitive conclusion regarding the formation of approximate state designs could be made. As the noise rate and/or time $t$ increases, the total number of errors accrued during the circuit increases, and the states in the ensemble become less pure, as evidenced in the bottom panels of Fig.~\ref{fig:design_dist}. This leads to bounds that become less tight, in particular the upper bound.
	
	These results demonstrate that even in the presence of noise and miscalibrations between the quantum and classically simulated states, one can make concrete inferences regarding the closeness of the projected ensemble realised in experiment to being a quantum state design.

	\section{Fundamental limitations \label{sec:Limitations}}
	
	So far, we have introduced an optimization-based approach to inferring properties of post-measurement quantum state ensembles, constructed explicit two-sided bounds for various quantities, and demonstrated an immediate application of our method for the verification of emergent quantum state designs.  Evidently, the method described in this paper gives us some degree of ability to learn quantities of interest in the context of measurement-induced dynamics, but we have already seen that in some cases there are unavoidable limitations in terms of what can be unambiguously inferred from experimental data. In this section, we discuss some of these limitations in detail, with the aim to more precisely characterise the boundary between properties of the post-measurement conditional quantum states that can or cannot be inferred from experiment.
	
	\subsection{Interpretation and consequences of Theorem \ref{thm:ExtremePoints} \label{subsec:MixedCons}}
	
	Theorem \ref{thm:ExtremePoints} and its corollary immediately tell us something regarding what can be inferred about certain quantities---most obviously those that measure how mixed the conditional states $\rho^Q_z$ are. The average global purity $\mathbbm{E}_z \Tr[(\rho^Q_z)^2]$ and the average global von Neumann entropy $\mathbbm{E}_z S(\rho^Q_z)$ are both examples of this; let us consider the former for concreteness. Suppose that the true ensemble $\mathcal{E}^Q$ features states that are appreciably mixed on average, $\mathbbm{E}_z \Tr[(\rho^Q_z)^2] = 1 - \delta$, with $\delta > 0$. Even in the best-case scenario, where we have perfect classical simulations, we cannot make the range of feasible values [Eqs.~(\ref{eq:FeasibleRange}, \ref{eq:FeasibleSet})] significantly narrower than $\delta$, since this range must include both the true value $(1-\delta)$ and the value $1 - O(e^{-H_{\rm min}(p)})$ implied by Theorem \ref{thm:ExtremePoints}. (Here $H_{\rm min}(p) = -\log \max_z p_z$ is the min-entropy of the distribution $p_z$, which typically scales linearly with the number of measurements.) So, if the ensemble being probed is not close to being pure, $\delta > 0$, then although we might hope to obtain a good lower bound for the averaged purity, we cannot obtain a good upper bound, and hence there will always be some uncertainty in our conclusions. We can hope to learn the purity to a good accuracy if the ensemble being measured is itself close to pure (something we wouldn't know in advance, but could verify using our bounds). Indeed, for perfect simulation, the bound \eqref{eq:PurityLBSuperoperator} tends to the true purity $1 - \delta$, and hence the range of feasible values \eqref{eq:FeasibleSet} has an optimal width $\delta$.
	
	It may seem counter-intuitive that there remains a ambiguity in the average purity even when the simulation being used is perfect. This stems from the difference between knowing that the simulation is perfect, versus having a perfect simulation but not knowing (or assuming) that it is so. To learn something definitive about the ensemble, we cannot make such an assumption, and this means that we may not be able to make sharp conclusions even for perfect classical simulation. Note also that this limitation is not intrinsic to the inference method we are proposing in this paper: it is a fundamental obstruction, in that whatever strategy we employ, we cannot rule out the possibility of a mostly-pure state ensemble.
	
\begin{figure}
		\centering
		\includegraphics[scale=1]{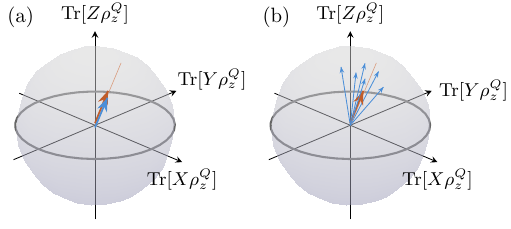}
		\caption{An illustrative example that demonstrates Theorem \ref{thm:Distinguish} implies a limitation our ability to sharply determine the purity of an ensemble. We consider a case where the quantum ensemble is given by the same mixed state for each $z$, i.e.~$\rho^Q_z = \rho^Q_0$, represented by a light blue arrow on the Bloch ball, with sub-unit length. (a) Perfect classical simulation also implies $\rho^C_z = \rho^Q_0$ (dark orange arrow). (b) When we measure quantum-classical correlations using $\rho^C_z = \rho^Q_0$, there is also another candidate ensemble made up of pure states which vary between each $z$ (multiple light blue arrows), distributed such that the same correlations are observed. Since we cannot \textit{a priori} know that our simulations are perfect (even if they are), we cannot use our experimental data to rule out the possibility that the true scenario is (b).}
		\label{fig:MixedExample}
	\end{figure}

	As a simple example that allows one to appreciate this idea intuitively, consider an ensemble defined for a single qubit ($d = 2$), where every conditional state is the same $\rho^Q_z = \rho_0^Q$, and this fixed state $\rho_0^Q$ is mixed. This is represented by the blue arrow on the Bloch sphere in Fig.~\ref{fig:MixedExample}(a). If our classical simulation were perfect, $\rho^C_z = \rho^Q_0$, then any quantum-classical correlator we measure corresponds to a $z$-independent choice of $A_z$ in Eq.~\eqref{eq:LinearObservableZ}, since the classical states are themselves $z$-independent. Once we measure some collection of quantum-classical correlators, we can consider which ensembles are consistent with these observations. The true ensemble is of course one possibility, but there are also alternative ensembles, made up of pure states which vary between different $z$, such that they average out to the same mixed state $\rho^Q_0$. This is shown by the many blue arrows in Fig.~\ref{fig:MixedExample}(b). Put simply, Theorem \ref{thm:ExtremePoints} reflects the fact that based on our experimental observations, we cannot determine whether the true states are mixed, or whether the states are pure but our classical simulations are inaccurate: Indeed, both of these have the effect of reducing the values of the quantum-classical correlators compared to the case of pure states with perfect simulation. To rule out the possibility of inaccurate classical simulation, one would have to estimate a number of quantities $R$ that itself scales with the cardinality of the ensemble $|\mathcal{Z}|$---effectively one for each possible value of $z$. This is of course not possible unless one uses a number of experimental repetitions $M$ that also scales with $|\mathcal{Z}|$.
	
	Looking at Figure \ref{fig:MixedExample}, we can roughly characterize the difference between panels (a) and (b) by saying that the mixed state in (a) is realised as a convex combination of the many different pure states in (b). When we consider more general averages of the form \eqref{eq:AverageProperty} with convex functions $G$ (beyond those like the average global purity), the ensembles (b) will have a higher value for this average compared to the true scenario (a), since taking convex combinations of states will reduce the value of $\bar{G}$. Therefore, the existence of consistent pure-state ensembles implied by Theorem \ref{thm:ExtremePoints} means that when the true ensemble is mixed, the upper bounds of convex functions cannot be made arbitrarily tight. As for lower bounds of convex averages, these can indeed be made tight as the fidelity of classical simulation is improved. Indeed, this is borne out in the various bounds we derived in Section \ref{sec:Constr}: Compare, e.g.~Eqs.~\eqref{eq:QuadLower} and \eqref{eq:QuadUpper} in the limit of $\rho^Q_z = \rho^C_z$. The lower bound tends to the true value, while the upper bound deviates from the true value by an amount $\|\mathcal{N}\|_\infty(1 - \mathbbm{E}_z\Tr[(\rho^Q_z)^2])$. One can see this asymmetry between the upper and lower bounds explicitly in Fig.~\ref{fig:noisy}.
	
	To conclude, the arguments presented above illustrate the conditions which must be satisfied if we wish to constrain the value of a particular nonlinear convex average \eqref{eq:AverageProperty} to within a small window: Not only should the classical simulation be accurate, but also the conditional states of the true ensemble should be close to pure---otherwise, it will be impossible to make the upper bound tight.\\
	
	\textit{Purification transition.---}One case where bounds on the global purity of the system are required is when probing purification dynamics \cite{Gullans2020}, and so this case deserves special attention. Here, an initially mixed state is subjected to a hybrid unitary-projective circuit, and one asks how the global purity increases over time. Since the states of interest are necessarily mixed even in the idealised case where the hybrid circuit is noiseless, one may conclude on the basis of Theorem \ref{thm:ExtremePoints} that purification dynamics cannot be probed experimentally. This is indeed the case if one actually prepares maximally mixed states as the inputs to the circuit; however if one instead uses a maximally entangled state between the system and a set of ancillas as the initial state, then the global state will be pure. The purity of the system can then be thought of as the purity of a subsystem of the global state, which can be bounded using the approach outlined in Section \ref{subsec:PurityUB}.
	
	Conveniently, using methods based on classical shadow tomography one can avoid having to use any ancillas in an actual experiment: By preparing randomized initial states and measuring the output states in random bases, it is possible to construct a classical shadow of the Choi state describing the hybrid dynamics. This Choi state is precisely the global state one would obtain by performing conventional shadow tomography on a purified system-plus-ancilla state. See, for instance, Refs.~\cite{Chen2020a,McGinley2022}.
	
	\subsection{Computational cost of classical simulation \label{subsec:Computation}}
	
	So far we have considered the classical states $\rho^C_z$ in quite general terms, without specifying how (and if) the states $\rho^C_z$ are constructed. Recall that the specific computational task required by our protocol is that upon obtaining an outcome $z$ in the experiment, we must compute and store a state $\rho^C_z$, which we construct on the basis of some model of how we expect the quantum device to behave. Whether or not this task is (even approximately) achievable is an intricate question, and has been discussed to some extent in other recent works where quantum-classical correlators have been introduced  \cite{Li2021a,Barratt2022,Lee2022,Li2023,Feng2023,Garratt2023,Hoke2023}. Here we explore some examples where we can answer this question in the affirmative or negative.
	
	Most obviously, some clear examples where such a classical simulation can be readily performed are in few body systems, or many-body systems where the full underlying dynamics is efficiently simulable, e.g.~for Clifford circuits or free fermionic systems. Going beyond these simple cases, the first issue to address is whether a representation of the simulated state can be stored using a scalable amount of classical memory (irrespective of whether it can be computed or not). One case where this is certainly possible is when the conditional states are defined on a $\mathcal{O}(1)$ number of qubits, even if the full quantum device is a many-body system. For example, in the projected ensemble \cite{Ho2022,Claeys2022,Cotler2023,Choi2023}, or teleportation transition \cite{Bao2024}, we can choose to measure many qubits, leaving only a few unmeasured. If, on the other hand, the post-measurement states consist of extensively many qubits, then the states themselves must have some appropriate structure which allows them to be represented using some variational ansatz, e.g.~a tensor network.
	
	In terms of the simulation strategy, the approach chosen will naturally depend on the specifics of the system, and so to make any further statements we must specialise to particular types of measurement-induced dynamics. One example where full simulation can be done efficiently is in the projected ensemble for a wavefunction generated by evolving a product state under some finite-time dynamics in one spatial dimension (see e.g.~Refs.~\cite{Ho2022,Ippoliti2022}). There, state of the unmeasured qubits conditioned on a particular measurement outcome can be constructed using a transfer matrix technique, the cost of which is linear in system size for any fixed time of evolution. Since the convergence towards emergent state designs is exponentially fast in time 1D \cite{Ippoliti2022}, this means that for the scheme described in Section \ref{subsec:designs}, the computational cost can remain small. The same goes for many-body teleportation protocols in one dimension \cite{Bao2024}. If the time of evolution were to scale with system size, or we move to higher dimensions, then efficient simulation may not be possible, although depending on the circuit being simulated there may still be viable options, see e.g.~Ref.~\cite{Napp2022}.
	
	For the case of one-dimensional hybrid quantum circuits, results on the measurement-induced phase transition indicate that the dynamics is efficiently simulable in the area-law phase using matrix product state-based techniques, but not in the volume law phase \cite{Skinner2019,Dehghani2023,Suzuki2023}. This suggests that experimentally learning properties of the post-measurement ensemble of conditional states without employing brute-force postselection is only possible in one phase. In the volume law phase, where simulation is presumably not possible, one cannot use the scheme described in this paper in practice, and we must instead ask whether anything nontrivial can be learned in the absence of a simulation---see Section \ref{subsec:NoSim}. Put simply, in these cases there exists a set of measurable properties \eqref{eq:LinearObservableZ} which could in principle be used to determine some ensemble property of interest, but to determine the correct operators $A_z$ is a computationally intractable task. See also Ref.~\cite{Garratt2023a}, where the question of whether entanglement can be probed in area vs.~volume law phases is considered.
	
	Beyond these examples, it is interesting to consider whether there may be scenarios in which the computational task required here (calculating conditional states for specific values of $z$) can be done efficiently, even if full simulation, which would also involve sampling from the probability distribution---a potentially hard task classically \cite{Terhal2004}---is not. One might also consider the possibility of using a second quantum device to perform the simulation itself: This certainly would be possible for the projected ensemble in 1D dual-unitary circuits by leveraging spacetime rotation \cite{Ippoliti2021}, but whether this can be done efficiently in other cases is unclear. Finally, we highlight that in settings where the dynamics features some particular structure, such as a continuous global symmetry, then it may be possible to make concrete inferences about the states in question by employing simulations that only use partial knowledge about the dynamics (e.g.~the distribution of gates employed rather than the exact gates chosen), which are computationally scalable. Indeed this was applied in Refs.~\cite{Barratt2022, Agrawal2023} to demonstrate the feasibility of probing charge-sharpening transitions in hybrid $\mathrm{U}(1)$-symmetric circuits. We leave these ideas to future work.
	
	\subsection{Can simulations be avoided? \label{subsec:NoSim}}
	
	Having highlighted the fact that constructing the simulated states $\rho^C_z$ may not always be possible, this raises the question of what can be learned without any advanced knowledge of the structure of the conditional states. To address this question in a sharp way, we consider a thought experiment, the implication of which is that if we cannot perform such a simulation, no information can be gained about the ensemble beyond the properties of the averaged state \eqref{eq:DMAverage} using a reasonable number of experimental repetitions $M$.
	
	Let us introduce the following hypothetical scenario: We are given access to a device (oracle) that when queried, outputs a label $z$ and state $\rho^Q_z$ sampled from an ensemble. The ensemble realised by the oracle is one of the following, chosen with equal \textit{a priori} probability
	\begin{subequations}
		\begin{align}
			\mathcal{E}_1^Q &= \{(p_z, \rho^Q_z)\}_{z \in \mathcal{Z}} \\
			\mathcal{E}_2^Q &= \{(p_z, \braket{\rho^Q})\}_{z \in \mathcal{Z}}.
		\end{align}
		\label{eq:EnsemblesDist}
	\end{subequations}
	where $\braket{\rho^Q}$ is the average state \eqref{eq:DMAverage}. After querying the oracle $M$ times, we are asked to determine whether the samples came from $\mathcal{E}_1^Q$ or $\mathcal{E}_2^Q$. We want to consider this problem because if we can't reliably distinguish these two ensembles, then we cannot hope to learn any properties of $\mathcal{E}_1^Q$ besides those that depend only on $\braket{\rho^Q}$.
	
	For the present purpose, we will broadly define a simulation as any method that allows us to obtain some (partial) prior knowledge of how the labels $z$ map onto properties of the individual states $\rho^Q_z$. Thus, in the scenario we are thinking about where we do not have access to a simulation, we can only employ strategies that treat each $z$ on an equal footing, i.e.~we cannot exploit of any contextual information about what the labels $z$ represent. To make this idea concrete, we define a simulation-free strategy as one that works the same if all labels $z$ output by the oracle are first permuted by some arbitrary $\tau \in \Sigma_{|\mathcal{Z}|}$. In Appendix \ref{app:Distinguish}, we prove the following
	\begin{theorem}
		\label{thm:Distinguish}
		Given coherent access to $M$ independent samples of quantum states from one of the two ensembles $\mathcal{E}_1^Q$ or $\mathcal{E}_2^Q$ [Eqs.~\eqref{eq:EnsemblesDist}], chosen with equal \textrm{a priori} probability, then any simulation-free strategy to distinguish $\mathcal{E}_1^Q$ from $\mathcal{E}_2^Q$ succeeds with probability no greater than
		\begin{align}
			p_{\rm succ} \leq \frac{1}{2} + M^2\sum_z p_z^2.
		\end{align}
	\end{theorem}
	Thus, in the regime where $M \lesssim 2^{H^{(2)}[p_z]}$, where $H^{(2)}[p_z] \coloneqq -\log_2(\sum_z p_z^2)$ is the collision entropy of the distribution $p_z$, which typically scales linearly with the number of measurements, we cannot reliably distinguish a given ensemble $\mathcal{E}_1^Q$ from one where the average state $\braket{\rho^Q}$ is supplied independently of $z$. As a consequence, we cannot hope to learn any property of an ensemble that is not contained within the average state \eqref{eq:DMAverage}, unless we obtain a number of samples that scales exponentially with the collision entropy of $p_z$. This is even the case when we can access all $M$ copies of the state simultaneously, which encompasses situations where we employ adaptive measurement strategies, i.e.~protocols where the gates and measurements that we apply can be chosen in a way that depends on measurement outcomes that occurred earlier. Hence, Theorem \ref{thm:Distinguish} establishes that the postselection problem can only be avoided if we incorporate some prior knowledge about the states $\rho^Q_z$ into our learning strategy.
	
	\section{Conclusion and outlook \label{sec:Concl}}
	
	In this work, we have introduced a scheme that allows one to infer properties of an ensemble of quantum states generated by dynamics that involve measurements. We avoid postselection by insisting that the quantities we directly measure from the experiment are `estimable properties' [Eq.~\eqref{eq:LinearObservableZ}], which can be computed using a scalable number of experimental repetitions. Then, information about an ensemble averaged property \eqref{eq:AverageProperty} can be indirectly inferred by solving an optimization problem, as defined in Section \ref{sec:ConvOptimization}. Our method gives one a lower and upper bound for the desired quantity, which can be made narrow by employing simulations of the quantum device on a classical computer. Crucially, the conclusions we make regarding the ensemble of states generated by the quantum device are not contingent on any \textit{a priori} assumptions about the accuracy of the simulation: Any bounds we construct are entirely rigorous.
	
	Our results clearly have immediate implications for near-term experiments that probe measurement-induced dynamics. In Section \ref{subsec:designs}, we saw how the method used here can be used to infer the emergence of quantum state designs in the projected ensemble. For other types of measurement-induced phenomena beyond this example, once the relevant order parameters and/or figures of merit are identified, and corresponding inequalities of the kind presented in Section \ref{sec:Constr} are derived, one can immediately start using these bounds in the spirit of this work to indirectly infer the value of this quantity realised in some experiment. In particular, by virtue of having post-measurement states defined on a small number of qubits, we anticipate that experiments for witnessing many-body teleportation transitions \cite{Bao2024,Hoke2023} should be ideal settings for the application of our method. 
	
	Thinking more generally, the arguments presented in Section \ref{sec:Limitations} point to a sharp distinction between scenarios where properties of the ensemble of post-measurement conditional states can or cannot be inferred from experimental data. One such condition is on the states being probed: If these are not close to being pure, then there will inevitably be some uncertainty in the value of the desired quantity. The other pertains to the feasibility of classical simulation: If simulation is not possible, then the states generated by the dynamics are indistinguishable from the case where the average state \eqref{eq:DMAverage} is realised every time. The implication is that if we hope to use the conditional states $\rho^Q_z$ as resources for some useful task, then we must necessarily have some prior knowledge about their structure, i.e.~some model to guide our expectation for how $\rho^Q_z$ depends on $z$. This basic expectation should be borne in mind when considering possible extrinsic applications of measurement-induced physics in future studies.
	
	While the approach we introduce here is designed with the aim to characterise properties of the post-measurement states, it is helpful to make comparisons to other scenarios where one wishes to make other kinds of inferences about processes that generate both quantum and classical data. One example that fits particularly well into this category is the problem of parameter estimation for continuously monitored few-body systems \cite{Mabuchi1996,Gambetta2001,Tsang2012,Gammelmark2013a,Gammelmark2014}, which has applications in quantum metrology. There, one again has to deal with the fact that every repetition of the experiment results in a different non-deterministic outcome, which can be handled by using a parallel classical simulation of the dynamics conditioned on those quantum trajectories that arise in the experiment. A key difference is that in parameter estimation, a particular starting assumption is made that the generator of dynamics is given by some model with a fixed number of unknown parameters, which we wish to infer based on the outcomes of the measurements we make. Because of this strong assumption, inferences can be made purely from the observed distribution of outcomes, without measuring state of the system at the end of the experiment. In cases where such model-based assumptions can be reliably applied, it may well be possible to incorporate these into our scheme through modification of the space $\mathcal{K}$, and this might perhaps allow one to overcome some of the limitations of Theorem \ref{thm:ExtremePoints}. In fact, in the even more structured setting where one wishes to learn the best candidate out of a set of hypotheses for a quantum-classical process, general bounds on the required sample complexity are known \cite{Caro2021,Fanizza2022}, and so insights from these works could suggest strategies for overcoming these limitations.
	
	One type of scenario that has not been explicitly considered in this work is where the outcome of measurements are used to determine some subsequent dynamics, i.e.~feedback is employed. In such scenarios, which include error correction \cite{Lidar2013} and measurement-based computation \cite{Raussendorf2001,Raussendorf2003} as special cases, nontrivial behaviour can arise even in the averaged state \cite{Piroli2021,Mcginley2022b,Iadecola2022,Friedman2023,Buchhold2022, Odea2022, Sierant2023}, and this can be probed in experiment using conventional learning techniques (though one should note that the physics of the averaged state is still distinct from those of the individual trajectories $\rho^Q_z$ \cite{Odea2022,Sierant2023,Lemaire2023}). Still, if one is interested in using such `interactive dynamics' to generate states with some desired property, then the choice of feedback applied could in principle be made based on some knowledge of the post-measurement states prior to the conditional unitary operation. The scheme introduced here can in principle be used to characterise these `pre-feedback' states, which can then be used to make a constructive choice of feedback algorithm. Indeed, this type of experimental flow arises in certain proposals of measurement-based computation \cite{Stephen2017}.
	
	\begin{acknowledgements}
		I am grateful to Sam Garratt and Wen Wei Ho for insightful discussions. Support from Trinity College, Cambridge is acknowledged.
	\end{acknowledgements}
	
	\appendix
	\section{Proof of Theorem \ref{thm:ExtremePoints} \label{app:Extreme}}
	
	To establish Theorem \ref{thm:ExtremePoints}, we will need the following result, which is a restatement of claim (5.8) in Ref.~\cite{Dubins1962}.
	\begin{lemma*}[\textbf{Dubins} \cite{Dubins1962}]
		Let $\mathcal{K}_0$ be a linearly bounded convex subset of a finite-dimensional vector space $\mathcal{M}$. Let $\mathcal{A}$ be an intersection of $R$ hyperplanes in $\mathcal{M}$, i.e.~a linear subspace of codimension $R$. Then the extreme points of the intersection $\mathcal{K} = \mathcal{K}_0 \cap \mathcal{A}$ are elements of the $R$-skeleton of $\mathcal{K}_0$, namely the union of all faces of $\mathcal{K}$ of dimension less than or equal to $R$.
	\end{lemma*}
	When applied to the feasible space $\mathcal{K}$ defined in Eq.~\eqref{eq:FeasibleSet}, we can reduce our problem to a study of the $R$-skeleton of $\mathcal{K}_0$. Recall that a face of a convex set $\mathcal{K}_0$ is a subset $\mathcal{F} \subset \mathcal{K}_0$ with the property that if $x \in \mathcal{F}$ is a convex combination of two other elements $x = \lambda x' + (1-\lambda)x''$, where $x', x'' \in \mathcal{K}_0$, then this implies that $x', x''  \in \mathcal{F}$.  All faces have a dimension, which is the dimension of the smallest affine set containing it---for example, the faces of dimension zero are singletons, each containing  an extreme points of $\mathcal{K}_0$. Here we are interested in the faces of dimension at most $R$.
	
	The set $\mathcal{K}_0$ in question is a Cartesian product of $|\mathcal{Z}|$ copies of the space of density matrices $\mathcal{D}$ over a Hilbert space of dimension $d$. Then, our first step is to show that the faces of $\mathcal{K}_0$ are products of faces of $\mathcal{D}$. To demonstrate this, consider a face $\mathcal{F} \subset \mathcal{C}$, where $\mathcal{C} = \mathcal{C}_1 \times \mathcal{C}_2$, with $\mathcal{C}_{1,2}$ arbitrary convex sets. For a given $x_1 \in \mathcal{C}_1$, denote $\mathcal{F}_2(x_1) \subset \mathcal{C}_2$ as the set of points $x_2$ such that the pair $(x_1, x_2)$ is an element of $\mathcal{F}$. Note that $\mathcal{F}_2(x_1)$ is a face of $\mathcal{C}_2$, since if $x_2 = \alpha y_2 + (1-\alpha)z_2$, then we have $(x_1, y_2) \in \mathcal{F}$ and $(x_1, z_2) \in \mathcal{F}$, which by definition implies $y_2, z_2 \in \mathcal{F}_2(x_1)$. Now take two points $x_1$, $x_1'$ for which $\mathcal{F}_2(x_1)$ is nonempty. By convexity of $\mathcal{F}$, if $x_2 \in \mathcal{F}_2(x_1)$ and $x_2' \in \mathcal{F}_2(x_1')$, then for any $\alpha \in [0,1]$ we have $(\alpha x_1 + (1-\alpha)x_1', \alpha x_2 + (1-\alpha)x_2' ) \in \mathcal{F}$. Then, by the definition of a face we have $x_2, x_2' \in \mathcal{F}_2(\alpha x_1 + (1-\alpha)x_1')$. This implies that $\mathcal{F}_2(x_1) = \mathcal{F}_2(x_1')$, and hence for any $x_1$, $\mathcal{F}_2(x_1)$ is either empty or equal to a fixed set $\mathcal{F}_2$, and we conclude that $\mathcal{F} = \mathcal{F}_1 \times \mathcal{F}_2$, where $\mathcal{F}_2$ is a face of $\mathcal{C}_2$. The same logic applied to the analogously defined set $\mathcal{F}_1(x_2)$ gives us that $\mathcal{F}_1$ must be a face of $\mathcal{C}_1$. If we apply this to the multiple product $\mathcal{K}_0 = \bigtimes_{z \in \mathcal{Z}} \mathcal{D}$, we conclude that faces of $\mathcal{K}_0$ are products of $|\mathcal{Z}|$ faces of $\mathcal{D}$, as claimed above.
	
	With this understood, note that the $n$-dimensional faces of $\mathcal{K}_0$ are products of faces of $\mathcal{D}$ whose dimensions sum to $n$. Thus, at least $(|\mathcal{Z}|-n)$ factors of any $n$-dimensional face of $\mathcal{K}_0$ will be zero-dimensional. Since a zero dimensional face is an extreme point, and the extreme points of $\mathcal{D}$ are pure states \cite{Bengtsson2013}, we have that any element of an $n$-dimensional face of $\mathcal{K}_0$ are ensembles with no more than $n$ non-pure states. We conclude that extreme points of $\mathcal{K}$, which according to the Lemma quoted above belong to the $R$-skeleton of $\mathcal{K}_0$, must have no more than $R$ non-pure states, thus competing the proof of Theorem \ref{thm:Distinguish}. \hfill $\square$

	The corollary quoted in the main text follows since $\mathcal{K}$ is a compact, convex subset of $\mathcal{M} = \bigoplus_{z \in \mathcal{Z}}\mathcal{B}(\mathcal{H})$, which is a finite-dimensional vector space, and hence the Krein-Milman theorem applies. Because $\mathcal{K}$ is necessarily non-empty, at least one extreme point of $\mathcal{K}$ must exist.
	
	\section{Derivations of lower and upper bounds \label{app:Bounds}}
	
	In Section \ref{sec:Constr}, we outlined how approximate solutions to the relevant optimization problems could be found for various specific quantities. Here we include some of the details of these derivations.

	\subsection{Quadratic observables \label{app:QuadLower}}
	
	Here we prove the lower bound \eqref{eq:QuadLower}, which is obtained using a similar line of reasoning to that discussed in Section \ref{subsec:PurityLB}. We start with the case where the superoperator $\mathcal{N} = |O\rrangle \llangle O|$, i.e.~we are considering the average of $G(\rho^Q_z) = \Tr[\rho^Q_z O]^2$. The case of general positive semi-definite $\mathcal{N} \succeq 0$ will follow quickly from this case.
	
	We can assume that $\Tr[O] = 0$ without loss of generality, since any component of $O$ proportional to $\mathbbm{I}$ can be subtracted off, yielding an additional term $\propto \Tr[\braket{\rho^Q}O]$, which is a property of the average state and hence can be measured directly. Then, the dual function $h_-(\lambda_i)$ corresponding to this particular primal problem takes the form
	\begin{align}
		\label{eq:H2MinF2}
		h_-(\lambda_i) &= \sum_z p_z F_{2,-}\left(O, {\textstyle \sum_i} \lambda_i \tilde{A}_z^{(i)}\right) + \sum_i \lambda_i(b_i - a_i)
	\end{align}
	where $a_i$, $\tilde{A}_z^{(i)}$ are as in Eq.~\eqref{eq:PurityMinF2}, and we have defined the function
	\begin{align}
		F_{2,-}(O, C) \coloneqq \inf_{\rho \in \mathcal{D}}\Big( \Tr[\rho O]^2 - \Tr[\rho C]\,\Big)
		\label{eq:OptimF2}
	\end{align}
	by analogy to Eq.~\eqref{eq:OptimF2Purity}.
	While we cannot exactly solve this problem analytically in full generality, it is possible in the case where $\mathcal{H}$ is a single qubit (Hilbert space dimension $d = 2$). Our solution to this case will prove instructive when it comes to treating the more general scenario $d \geq 2$ later; we thus briefly specialise in the following.\\
	
	\textit{Single qubit.---}When we fix $d = 2$, since both $C$ and $O$ are traceless, one can rescale and perform rotations in Hilbert space appropriately to map the problem of evaluating $F_{2,-}(O,C)$ to the case $O = Z$, and $C = \alpha_1 X + \alpha_3 Z$, where $(X, Y, Z)$ are Pauli matrices, and $\alpha_1, \alpha_3 \geq 0$ are nonnegative scalars. Evidently, $\alpha_1$ and $\alpha_3$ are the components of $C$ orthogonal and parallel to $O$, respectively.  This effectively reduces $F_{2,-}$ to a function of two variables $(\alpha_1, \alpha_3)$.
	
	Another useful aspect of the qubit problem that helps us here is that the the space of density matrices $\mathcal{D}$ has a simple geometry that can be straightforwardly characterized: The conditions that $\rho$ is Hermitian and satisfies $\Tr[\rho] = 1$, $\Tr[\rho^2] \leq 1$ are necessary and sufficient for $\rho$ to be a valid density matrix. This leads to the notion of the Bloch ball, which is a helpful visualization of the space $\mathcal{D}$: any density matrix can be written as $|\rho\rrangle = (|\mathbbm{I}\rrangle + n_1|X\rrangle + n_2|Y\rrangle + n_3|Z\rrangle)/2$, with $\vec{n} = (n_1, n_2, n_3)$ a 3-dimensional vector that specifies the state, which belongs to the unit ball $|\vec{n}| \leq 1$. Since the function to be extremised is independent of $n_2$, we can reduce this to
	\begin{align}
		\label{eq:F2DefQubit}
		F_{2,-}^{(d=2)}(\alpha_1, \alpha_3) = \inf_{n_1^2+n_3^2 \leq 1} \Big[n_3^2 - \alpha_1n_1 - \alpha_3n_3 \Big]
	\end{align}
	A solution to this problem can be formally written down in terms of the roots of a certain polynomial equation, but the resulting expression is rather cumbersome and not particularly informative. Rather, it is useful to consider the behaviour of the above function in the vicinity of $\alpha_1 = 0$. By considering small perturbations around the $\alpha_1 = 0$ solution, we obtain the expansion
	\begin{align}
		F_{2,-}^{(d=2)}(\alpha_1, \alpha_3) &= \begin{dcases} -\frac{\alpha_3^2}{4} - |\alpha_1|\sqrt{1 - \frac{\alpha_3^2}{4}}   & |\alpha_3| \leq 2 \\
			1 - |\alpha_3|  & |\alpha_3| > 2
		\end{dcases}\nonumber\\
		&+ \mathcal{O}(\alpha_1^2)
		\label{eq:F2SeriesQubit}
	\end{align}
	which is achieved at
	\begin{align}
		n_3 = \mathop{\text{sgn}}(\alpha_3) \times \min(1, |\alpha_3|/2) + \mathcal{O}(\alpha_1).
		\label{eq:F2OptimalQubit}
	\end{align}
	Looking at Eq.~\eqref{eq:F2SeriesQubit}, one notices that the behaviour of $F_{2,-}^{(d=2)}$ in the directions $\alpha_1$ and $\alpha_3$ is markedly different near the point $\alpha_{1,3} = 0$: The function decreases linearly along $\alpha_1$, and quadratically along $\alpha_3$. Bearing in mind that $\alpha_{1,3}$ are linear combinations of the Lagrange multipliers $\lambda_i$, we consider a stability analysis of the function $h_{-}(\lambda_i)$, which is to be maximized, around the candidate point $\lambda_i = 0$. Qualitatively speaking, we will find that it is favourable to increase $\lambda_i$ away from zero for those $i$ where the operators $A_z^{(i)}$ are predominately along the $Z$ direction in operator space ($\alpha_3$ dominates), but not those along orthogonal directions ($\alpha_1$ dominates). This is because for the former subset of $\lambda_i$, the second term in Eq.~\eqref{eq:H2MinF2} will increase faster than the first term decreases as $\lambda_i$ is varied away from zero.
	
	This rough intuition can be made concrete most straightforwardly if we assume that for each $i$, $A^{(i)}_z$ is either orthogonal to or parallel to $Z$. (Such an assumption is not particularly restrictive---if $A^{(i)}_z$ are measurable, then we can always decompose $A^{(i)}_z = \sum_\mu a_{i \mu z} \sigma^\mu$ in terms of components along some basis of operators $\sigma^\mu$, which we are free to choose such that one of the $\sigma^\mu$ is proportional to $O$. Then, using our shadow-based scheme we can always choose to measure the enlarged set of operators $A^{(i,\mu)}_z \coloneqq a_{i\mu z}\sigma^\mu$ for each pair $(i,\mu)$ separately without losing any information, simply by altering the classical post-processing.) Let us denote the set of $i$ for which $\llangle A_z^{(i)}|Z\rrangle = 0$ as $\mathcal{I}_{\perp}$, and those for which $A_z^{(i)} = c_{z i}Z$ as $\mathcal{I}_\parallel$. We temporarily fix $\lambda_{i \in \mathcal{I}_\perp} = 0$ by hand and then optimize over $\lambda_{i \in \mathcal{I}_{\parallel}}$. Using the expansion Eq.~\eqref{eq:F2SeriesQubit} with the appropriate replacement of parameters, we obtain
	\begin{align}
		h_-(\lambda_i)|_{\lambda_i \in \mathcal{I}_\perp = 0} &= \sum_z p_z F_{2,-}^{(d=2)}\Bigg(0, \sum_{i\in \mathcal{I}_\parallel} \lambda_i c_{z i}\Bigg) \nonumber\\ &+\sum_{i \in \mathcal{I}_\parallel} \lambda_ib_i.
	\end{align}
	(Note that we have $a_i = 0$ for $i \in \mathcal{I}_\parallel$.) From this point, our arguments follow a similar structure to those of Subsection \ref{subsec:PurityLB}. We can use the bound $F_{2,-}^{(d=2)}(0, \alpha_3) \geq -\alpha_3^2/4$ to obtain an analogous expression to Eq.~\eqref{eq:PurityMinLambda}
	\begin{align}
		h_-(\lambda_i)|_{\lambda_{i \in \mathcal{I}_\perp}=0} \geq -\frac{1}{4}\sum_{i,j\in\mathcal{I}_{\parallel}}\lambda_i J_{ij} \lambda_j + \sum_{i\in\mathcal{I}_{\parallel}} b_i \lambda_i,
		\label{eq:HMinLambda}
	\end{align}
	where we have introduced $J_{ij}$, an analogue of $L_{ij}$ in Eq.~\eqref{eq:PurityMinimumL}
	\begin{align}
		J_{ij} = \sum_z p_z c_{z i}c_{z j}.
	\end{align}
	(Eq.~\eqref{eq:HMinLambda} actually becomes an equality if we have $|\sum_{i\in \mathcal{I}_\parallel} \lambda_i c_{z i}| \leq 2$ for all $z$.) As before, this can be maximized to obtain a certificate
	\begin{align}
		\max_{\lambda_{i}\, :\, i \in \mathcal{I}_\parallel}h_-(\lambda_i) \geq \sum_{i,j \in \mathcal{I}_\parallel} b_i [L^{-1}]_{ij}b_j \eqqcolon \tilde{h}_-
		\label{eq:DualMaximumParallel}
	\end{align}
	Again, optimality of this candidate solution to the dual problem is not guaranteed, i.e.~$\tilde{h}_-$ might be strictly smaller than the true optimal value $g_-^*$. Even so, the above can be evaluated straightforwardly using experimental data, and can be used as a lower bound for the quantum-quantum correlator $\mathbbm{E}_z \Tr[\rho_z^Q O]^2$.

	\textit{Beyond a single qubit.---}When we generalize beyond the case of a single qubit $d = 2$, the optimization over density matrices in higher dimensional Hilbert spaces is made more complicated by the geometric structure of the space $\mathcal{D}$, which does not have simple interpretations such as the Bloch ball. For instance, expressions analogous to Eq.~\eqref{eq:F2DefQubit} are not so straightforward. Nevertheless, we can carry the intuition gained above forward to this case to anticipate that reasonable bounds can be found by optimizing only $\lambda_i$ for $i \in \mathcal{I}_\parallel$, where now $\mathcal{I}_{\parallel}$ denotes the set of $i$s for which $A^{(i)}_z = c_{zi}O$ for scalars $c_{iz}$, whose complement $\mathcal{I}_{\perp}$ is made up of operators orthogonal to $O$ in operator space, namely $\Tr[A_z^{(i)} O] = 0$ for $i \in \mathcal{I}_\perp$.
	
	Following the same logic as before, we set $\lambda_i = 0$ for $i \in \mathcal{I}_\perp$, after which the second argument of the function $F_{2,-}$ in Eq.~\eqref{eq:H2MinF2} becomes proportional to $O$. The resulting minimization problem \eqref{eq:OptimF2} can then be directly evaluated for arbitrary $d$
	\begin{align}
		F_{2,-}(O, \alpha O) = -\|O\|_\infty^2 \times \min(\alpha^2/4, |\alpha|)
	\end{align}
	where $\|O\|_\infty$ denotes the spectral norm, equal to the largest singular value of $O$. This solution has the same structure as the qubit case \eqref{eq:F2SeriesQubit}, and hence the same logic can be used as before to reproduce the bound Eq.~\eqref{eq:DualMaximumParallel}. That is, the same expression \eqref{eq:DualMaximumParallel} for $\tilde{h}_-$ as a lower bound for $g_-^*$ for arbitrary $d$.
	
	Now, following the same approach as in Section \ref{subsec:PurityLB}, we can use quantum-classical correlators to choose the operators $A_z^{(i)}$. The simplest case, where we use the standard quantum-classical correlator \eqref{eq:CorrQC} as the only constraint, yields
	\begin{align}
		\braket{O\otimes O}^{QQ} \geq \frac{\big[\!\braket{O\otimes O}^{QC}\big]^2}{\braket{O\otimes O}^{CC}},
		\label{eq:IneqCauchySchwartz}
	\end{align}
	which can be interpreted as a resuly of the Cauchy-Schwartz inequality, or equivalently the standard inequality $\Var(X)\Var(Y) \geq \Cov(X,Y)^2$ for classical random variables $X$, $Y$. If we make use of quantum-classical correlators beyond just $\braket{O \otimes O}^{QC}$, then we can in principle use all the information contained in the superoperators $\eta^{QC}$, $\eta^{CC}$ [Eq.~\eqref{eq:CQSuperoperators}]. This allows us to improve the simple bound above to
	\begin{align}
		\braket{O\otimes O}^{QQ} \geq \llangle O|  \eta^{QC} \zeta^{CQ}|O  \rrangle
		\label{eq:SqMinSuperoperator}
	\end{align}
	where $\zeta^{CQ}$ is the superoperator implicitly defined in \eqref{eq:ZetaQCImplicit}.
	
	Now, for more general convex quadratic observables ($\mathcal{N} \succeq 0$), we can always perform a decomposition of the superoperator $\mathcal{N} = \sum_a \mu_a |C_a \rrangle \llangle C_a|$, where $\mu_a \geq 0$ are eigenvalues and $C_a$ are eigenoperators. Due to the non-negativity of $\mu_a$, we can apply the bound \eqref{eq:SqMinSuperoperator} with $O = C_a$ for each $a$ separately, which gives Eq.~\eqref{eq:QuadLower}.
	
	\subsection{Numerically stable upper bound for von Neumann entropy \label{app:vNUpper}}
	
	In the main text, we derived a general upper bound for the average von Neumann entropy \eqref{eq:vNMaxLambda}, which for the special case $A_z = -\log \rho^C_z$ reduces to the quantum-classical entropy introduced in Ref.~\cite{Garratt2023a}. Unfortunately, this leads to an expression that is numerically unstable when the classical states $\rho^C_z$ are near-singular, due to the need to take the logarithm of the operators $\rho^C_z$. However, thanks to the flexibility of our optimization-based approach, we can devise a simple solution, where we modify the chosen operators $A_z$ to obtain a better bound. Here we consider the case where we measure the two observables defined in Eq.~\eqref{eq:vNRegularizedObservables}, denoted $A^{(1,2)}_z$. For each of these two observables we have a Lagrange multiplier $\lambda_{1,2}$. In terms of the expectation values $\braket{A^{(1,2)}_z}$, the dual function can be obtained by minimizing with respect to $\rho^Q_z$, giving
	\begin{align}
		h_+(\lambda_{1,2}) &= \mathbbm{E}_z\bigg[\log\Big(\Tr[(\Pi^>_z \rho^C_z \Pi^>_z)^{\lambda_1}] + e^{-\lambda_2}\text{rank}(\Pi_z^\leq)\Big) \nonumber\\
		&+ \lambda_1 \braket{A^{(1)}_z} + \lambda_2  \braket{A^{(2)}_z}\bigg]
	\end{align}
	Optimizing for $\lambda_{1,2}$ cannot be done exactly using analytical methods; however, based on the same considerations as for Eq.~\eqref{eq:vNMaxDualFunction}, it is reasonable to set $\lambda_1 = 1$ in the above, after which the minimization over the remaining Lagrange parameter $\lambda_2$ can be done approximately. In the limit where $\overline{\delta^Q} \coloneqq \braket{A_z^{(2)}} = \mathbbm{E}_z \Tr[\rho^Q_z \Pi^\leq_z]$ is small (i.e.~the overlap of the quantum states with the near-singular eigenstates of $\rho^C_z$ is small on average), we find that the value $\lambda_2 = \log(\overline{r}(1-\overline{\delta^Q})/\overline{\delta^Q}(1-\overline{\delta^C}))$ is approximately optimal, where we use the shorthand $\overline{r} = \mathbbm{E}_z r_z = \mathbbm{E}_z \text{rank}[\Pi_z^\leq]$ and $\overline{\delta^C} = \mathbbm{E}_z \delta^C_z = \mathbbm{E}_z \Tr[\rho^C_z \Pi^\leq_z]$. This gives us a regularized bound
	\begin{align}
		&\mathbbm{E}_z S_{\rm vN}(\rho^Q_z) \leq \sum_z p_z\Bigg[\!-\braket{A^{(1)}_z} \nonumber\\
		+& \log\left(1-\delta^C_z+\frac{\overline{\delta^Q}(1-\overline{\delta^C})}{(1-\overline{\delta^Q})}\frac{r_z}{\overline{r}}\right) - \overline{\delta^Q}\log\left(\frac{\overline{\delta^Q}(1-\overline{\delta^C})}{(1-\overline{\delta^Q})}\right)\Bigg].
	\end{align}
	The above simplifies considerably when the classical states are pure, where we have $\log'\rho^C_z = 0$, $r_z = (d-1)$, and $\delta^C_z = 0$, whence the result given in Eq.~\eqref{eq:vNUpperPure}.
	
	\subsection{Lower bound for subsystem von Neumann entropy \label{app:vNLower}}
	
	In the main text, we showed that the minimization problem for the average von Neumann entropy of a subsystem $Q_1$ could be turned into a convex optimization problem by way of the conditional entropy \eqref{eq:CondEntropyDef}, which is concave, and hence can be maximized. To make progress, it will be useful to introduce a variational characterization of the conditional quantum entropy \cite{Tomamichel2014}
	\begin{align}
		S(\rho^Q|Q_1) = \sup_{\sigma^{Q_1} \in \mathcal{D}_1}\Big[-S(\rho^Q \| \sigma^{Q_1}\otimes \mathbbm{I}_{Q_2})\Big],
		\label{eq:CondEntropyVariational}
	\end{align}
	where $S(\rho \| \sigma) = \Tr[\rho \log \rho] - \Tr[\rho \log \sigma]$ is the quantum relative entropy. After constructing the dual function for maximizing the conditional entropy, we can use the above representation, after which the result \eqref{eq:vNMaxSubProblem} can be used to perform the optimization over $\rho^Q_z$, thus giving
	\begin{align}
		&\max_{\blockstyle{\rho}\in \mathcal{K}}\Big[\mathbbm{E}_z S(Q_2|Q_1)_{\rho^Q_z}\Big] = \min_{\lambda_i}\bigg( {\textstyle\sum_i} \lambda_i b_i \nonumber\\ +&  \max_{\sigma^{Q_1}_z \in \mathcal{D}_1} \mathbbm{E}_z \log \Tr\bigg[\exp\Big(\log \sigma^{Q_1}_z\otimes \mathbbm{I}^{Q_2}-{\textstyle \sum_i} \lambda_i A_z^{(i)}\Big)\bigg]\bigg)
	\end{align}
	Due to the form of Eq.~\eqref{eq:CondEntropyVariational}, we have an extra maximization step to perform over the states $\sigma^{Q_1}_z$. This cannot be solved exactly in full generality, but we can use the Golden-Thompson inequality $\Tr[e^{A+B}] \leq \Tr[e^A e^B]$ for Hermitian matrices $A$, $B$ to obtain the bound \eqref{eq:vNMinInfNorm}.
	
	\section{Proof of Theorem \ref{thm:Distinguish} \label{app:Distinguish}}
	
	Here we provide a proof of Theorem \ref{thm:Distinguish}, which is a statement regarding our ability to distinguish between the two ensembles \eqref{eq:EnsemblesDist}. First, we note that having coherent access to $M$ samples from the ensembles $\mathcal{E}^Q_1$, $\mathcal{E}^Q_2$ is equivalent to owning a single copy of the following respective quantum-classical states
	\begin{subequations}
		\begin{align}
			\rho_{1,M} &= \left(\sum_z p_z \ket{z}\bra{z} \otimes \rho^Q_z\right)^{\otimes M} \\
			\rho_{2,M} &= \left(\sum_z p_z \ket{z}\bra{z} \otimes \braket{\rho^Q}\right)^{\otimes M}
		\end{align}
		\label{eq:QCStatesDist}
	\end{subequations}
	In this representation, the labels are stored in $M$ classical registers each prepared in a state $\ket{z}\bra{z}$, and the conditional states are simultaneously stored in $M$ separate quantum registers. The problem of deciding which of the two scenarios \eqref{eq:QCStatesDist} is realised by the oracle is a form of hypothesis testing, which is a well-studied problem. Without any constraints on our hypothesis testing strategy, a standard result due to Helstrom \cite{Helstrom1969} says that the best strategy is equivalent to performing a generalized measurement described by a two-component POVM $E_1 + E_2 = \mathbbm{I}$, where $E_1$ is the projector onto the space of positive eigenvalues of $\rho_{1,M} - \rho_{2,M}$. When one employs this strategy, one can successfully determine which of the ensembles is realised with probability $p_{\rm succ} = (1/2) + \|\rho_{1,M} - \rho_{2,M}\|_1/4$, where $\|\rho-\sigma\|_1$ is the trace distance between states $\rho$, $\sigma$. As the number of copies $M$ increases, the success probability tends towards unity exponentially quickly \cite{Ogawa2004,Audenaert2007}, and so there always exists a strategy that allows us to distinguish between the ensembles \eqref{eq:EnsemblesDist} reliably.
	
	However, even though such an optimal hypothesis testing strategy may exist, an observer with limited computational power may not be able to determine what the correct measurement procedure should be. That is, although the ensembles may be distinguishable in an information-theoretic sense, they may not be computationally distinguishable. Since the question of the computational complexity of finding the optimal strategy will depend on the specifics of the ensemble $\mathcal{E}^Q$ in question, our starting point will 
	be to assume that simulations of the device of any kind are not available. This motivates our definition of a simulation-free strategy as defined in the main text.
	
	In this context, any simulation-free strategy can still be represented as a two-component POVM $E_{1,2}$ acting jointly on the $M$ classical and $M$ quantum registers which store the states \eqref{eq:QCStatesDist}, in the same way above---this encompasses protocols that feature measurements in entangled bases, adaptive strategies, and stochastic operations. Regardless, we want to restrict ourselves to operations that are symmetric among the different labels $z$, i.e.~we treat all labels equivalently.  This implies that we should make a restriction
	\begin{align}
		(\pi_\tau^{\otimes M} \otimes \mathbbm{I}_Q) E_1 (\pi_\tau^{\otimes M} \otimes \mathbbm{I}_Q) &= E_1 & \forall \tau \in \Sigma_{|\mathcal{Z}|}
		\label{eq:POVMSym}
	\end{align}
	where $\pi_\tau \ket{z} = \ket{\tau(z)}$ is a permutation operator on a single classical register, and $\mathbbm{I}_Q$ is the identity operator on all $M$ quantum registers. Making this restriction forbids us from using some prior knowledge of how the ensemble states $\rho^Q_z$ depend on the states $z$, i.e.~a means to simulate the dynamics.
	
	Whilst ensuring that Eq.~\eqref{eq:POVMSym} holds, we want to maximize the success probability $p_{\rm succ} = (1/2) + \Tr[E_1(\rho_{1,M} - \rho_{2,M})]/2$, as in standard hypothesis testing. Evidently, for any such operator $E_1$ we have $\Tr[E_1\sigma] = \Tr[E_1(\mathcal{S}\otimes \text{id}_Q)[\sigma]]$, where $\text{id}_Q$ is the identity superoperator on the quantum register, and
	\begin{align}
		\mathcal{S}[\sigma^C] \coloneqq \frac{1}{|\mathcal{Z}|!}\sum_{\tau \in \Sigma_{|\mathcal{Z}|}}\pi_\tau^{\otimes M} \sigma^C \pi_\tau^{\otimes M}
	\end{align}
	projects the operator $\sigma^C$ corresponding to the classical registers onto the subspace that is invariant under permutations of the labels $z$. Using this construction, it is straightforward to show that
	\begin{align}
		p_{\rm succ} = \frac{1}{2} + \frac{1}{2}\big\|(\mathcal{S}\otimes \text{id}_Q)[\rho_{1,M} - \rho_{2,M}]\big\|_1.
		\label{eq:PSuccTrace}
	\end{align}
	Now, we can write the argument of the trace norm as
	\begin{align}
		(\mathcal{S}\otimes \text{id}_Q)[\rho_{1,M} - \rho_{2,M}] &=  \sum_{\vec{z}} p(\vec{z}) \mathcal{S}[\ket{\tau(\vec{z})}\bra{\tau(\vec{z})}] \nonumber\\ & \otimes \left(\bigotimes_{i = 1}^M \rho^Q_{z_i} - \braket{\rho^Q}^{\otimes M}\right)
	\end{align}
	where we use the shorthand $p(\vec{z}) = \prod_{i=1}^M p_{z_i}$ and $\ket{\tau(\vec{z})} = \ket{\tau(z_1) \otimes \cdots \otimes \tau(z_M)}$. Now, for a string of labels $\vec{z}$, define the matrix $R_{ij}(\vec{z}) = \delta_{z_i, z_j}$, which specifies which pairs $(z_i, z_j)$ are equal for the given string $\vec{z}$. We have that $\mathcal{S}[\ket{\vec{z}}\bra{\vec{z}}]$ is equal to $\mathcal{S}[\ket{\vec{z}\,'}\bra{\vec{z}\,'}]$ if $R_{ij}(\vec{z}) = R_{ij}(\vec{z}\,')$ for all $i, j = 1, \ldots, M$, and they are orthogonal otherwise. The trace norm in Eq.~\eqref{eq:PSuccTrace} then becomes a sum over contributions from each possible value of $R_{ij}$
	\begin{widetext}
		\begin{align}
			\big\|(\mathcal{S}\otimes \text{id}_Q)[\rho_{1,M} - \rho_{2,M}]\big\|_1 = \sum_{R_{ij}\in \{0,1\}^{\times M^2}} \left\| \sum_{\vec{z} \in R_{ij}} p(\vec{z}) \left(\bigotimes_{i = 1}^M \rho^Q_{z_i} - \braket{\rho^Q}^{\otimes M}\right) \right\|_1
		\end{align}
	\end{widetext}
	where the notation $\vec{z} \in R_{ij}$ denotes a sum over all strings $\vec{z}$ for which $R_{ij}(\vec{z}) = R_{ij}$. At this point, we can start to apply some upper bounds. We separate out the term in the sum for which $R_{ij} = 0$, and we evidently have
	\begin{align}
		&\big\|(\mathcal{S}\otimes \text{id}_Q)[\rho_{1,M} - \rho_{2,M}]\big\|_1 \leq \sum_{z \notin 0} p(\vec{z}) \nonumber\\ +& \left\| \sum_{\vec{z} \in 0} p(\vec{z}) \left(\bigotimes_{i = 1}^M \rho^Q_{z_i} - \braket{\rho^Q}^{\otimes M}\right) \right\|_1
	\end{align}
	Now we use the definition of the average state, which implies that $\sum_{\vec{z}}p(\vec{z}) (\bigotimes_{i=1}^M \rho^Q_{z_i}) = \braket{\rho^Q}^{\otimes M}$. Hence, we can infer
	\begin{align}
		&\big\|(\mathcal{S}\otimes \text{id}_Q)[\rho_{1,M} - \rho_{2,M}]\big\|_1 \leq \sum_{z \notin 0} p(\vec{z}) \nonumber\\ +&
		\left\| \sum_{\vec{z} \notin 0} p(\vec{z}) \bigotimes_{i = 1}^M \rho^Q_{z_i} \right\|_1
	\end{align}
	from which we obtain
	\begin{align}
		p_{\rm succ} \leq \frac{1}{2} + \sum_{\vec{z} \in 0} p(\vec{z})
	\end{align}
	The second term in the above corresponds to the probability of at least one pair of labels $(z_i, z_j)$ being equal. A simple upper bound for this probability is ${M \choose 2}\sum_z p_z^2$ \cite{Mase1992}, and hence we obtain the bound quoted in Theorem \ref{thm:Distinguish}.

	\bibliography{verifiable_learning}
	
\end{document}